\documentclass[letterpaper,twocolumn,10pt]{article}
\usepackage{usenix2019_v3} % For I have become death; destroyer of style files
\usepackage{amssymb}
\usepackage{amsmath}

\usepackage{tikz,graphicx,balance,multirow,xspace,subcaption,longtable}
\usepackage{makecell,amsmath,cleveref,rotating}
\usepackage[T1]{fontenc}
\usepackage{enumitem}
\usepackage[draft,inline,nomargin,index]{fixme}
\usepackage{booktabs}
\usepackage{lastpage}
\usepackage{titling}

\usepackage[font={small}, textfont=md, skip=.2pt]{caption} % the commented version of this above is the original 
\newcommand{\tabularscale}{0.85}

\PassOptionsToPackage{hyphens}{url}
\usepackage{hyperref}
\newcolumntype{P}[1]{>{\arraybackslash}p{#1}}
\newcolumntype{X}[1]{>{\centering\arraybackslash}p{#1}}

\expandafter\def\expandafter\UrlBreaks\expandafter{\UrlBreaks%  save the current one
  \do\a\do\b\do\c\do\d\do\e\do\f\do\g\do\h\do\i\do\j%
  \do\k\do\l\do\m\do\n\do\o\do\p\do\q\do\r\do\s\do\t%
  \do\u\do\v\do\w\do\x\do\y\do\z\do\A\do\B\do\C\do\D%
  \do\E\do\F\do\G\do\H\do\I\do\J\do\K\do\L\do\M\do\N%
  \do\O\do\P\do\Q\do\R\do\S\do\T\do\U\do\V\do\W\do\X%
  \do\Y\do\Z}

\crefformat{section}{\S#2#1#3}
\crefformat{subsection}{\S#2#1#3}
\crefformat{subsubsection}{\S#2#1#3}

\fxsetup{theme=colorsig,mode=multiuser,inlineface=\itshape,envface=\itshape}
\FXRegisterAuthor{ht}{aht}{Hammas}
\FXRegisterAuthor{rn}{arn}{Rishab}
\FXRegisterAuthor{rs}{ars}{Rachee}
\FXRegisterAuthor{pp}{app}{Paul}

\hypersetup{%
  pdftitle={Glowing in the dark},
  pdfauthor={},
  pdfkeywords={},
  bookmarksnumbered,
  pdfstartview={FitH},
  colorlinks,
  citecolor=black,
  filecolor=black,
  linkcolor=black,
  urlcolor=blue,
  breaklinks=true,
  hypertexnames=false
}

\newcommand\clearrow{\global\let\rowmac\relax}

\newcommand{\para}[1]{{\vspace{0in} \bf \noindent #1 }}
\newcommand{\parait}[1]{{\vspace{0in} \em \noindent #1 }}

\newcommand{\subnet}[1]{\texttt{/#1}}

\newcommand{\Cg}{$\mathcal{C}$\xspace}
\newcommand{\Tg}[1]{$\mathcal{T}_{{\text{#1}}}$\xspace}
\newcommand{\rate}{$\mathcal{\mu}$\xspace}
\newcommand{\ddiff}[1]{$\Delta_{#1}^{\text{diff}}$\xspace}
\newcommand{\dcontrol}[1]{$\Delta_{#1}^{\text{\Cg}}$\xspace}
\newcommand{\meanscansperday}[2]{$\text{\rate}^{({#1})}_{\text{{#2}}}$\xspace}

\renewcommand{\footnotesize}{\scriptsize}

\newcommand{\eg}{e.g.,\ }
\newcommand{\etal}{et al.\xspace}
\newcommand{\ie}{i.e.,\ }

\newcommand{\nx}{{NXDOMAIN}\xspace}

\newcommand{\plb}{$P_{l}$\xspace}
\newcommand{\prb}{$P_{r}$\xspace}
\newcommand{\slb}{$S_{l}$\xspace}
\newcommand{\srb}{$S_{r}$\xspace}

\newcommand{\lb}{\textit{lower-byte}\ }
\newcommand{\rnd}{\textit{random}\ }

\usepackage{authblk}

\begin{document}

\pagestyle{plain} % removes running headers

\renewcommand{\sectionautorefname}{\S}
\renewcommand{\subsectionautorefname}{\S}
\renewcommand{\subsubsectionautorefname}{\S}
\date{}

\setlength{\droptitle}{-6em}   % Eliminate vertical space above title
\posttitle{\par\end{center}}   % tighten space between title and author block
\title{\bf Glowing in the Dark \\
\large{Uncovering IPv6 Address Discovery and Scanning Strategies in the Wild}}
% \author[1]{Hammas Bin Tanveer}
% \author[2]{Rachee Singh}
% \author[3]{Paul Pearce}
% \author[1]{Rishab Nithayanand}
% \affil[1]{The University of Iowa}
% \affil[2]{Microsoft Research}
% \affil[3]{Georgia Tech}
% \author[1]{Alice Smith}
% \author[2]{Bob Jones}
% \affil[1]{Department of Mathematics, University X}
% \affil[2]{Department of Biology, University Y}

% \author{%
% Hammas Bin Tanveer}
% \affil{University of Iowa}%
% \author{Rachee Singh}
% \affil{University of Iowa}%
\renewcommand\Authsep{, }
\renewcommand\Authand{}
\renewcommand\Authands{, }
\makeatletter
\renewcommand\AB@affilsepx{, \protect\Affilfont}
\makeatother
\setlength{\affilsep}{0.5em}
\author[1]{Hammas Bin Tanveer}
\author[2,3]{Rachee Singh}
\author[4]{Paul Pearce}
\author[1]{Rishab Nithyanand}
\affil[1]{University of Iowa}
\affil[2]{Microsoft}
\affil[3]{Cornell University}
\affil[4]{Georgia Tech}

% \author{[Paper \#XX, \pageref{LastPage} pages (including references)]}
% \author{Paper \#233}
% \author{Hammas Bin Tanveer}
% \affiliation{\institution{University of Iowa}}
% \author{Rachee Singh}
% \affiliation{\institution{Microsoft Research}}
% \author{Paul Pearce}
% \affiliation{\institution{Georgia Tech}}
% \author{Rishab Nithyanand}
% \affiliation{\institution{University of Iowa}}
% \renewcommand{\shortauthors}{H. Tanveer \etal}

% \author{
% {\rm Hammas Bin Tanveer}\\
% The University of Iowa
% \and
% {\rm Rachee Singh}\\
% Microsoft Research
% \and
% {\rm Paul Pearce}\\
% Georgia Tech
% \and
% {\rm Rishab Nithyanand}\\
% The University of Iowa 
% } % end author

\maketitle

\section*{Abstract}
% What is the problem?
In this work we identify scanning strategies of IPv6 scanners on the Internet.
% How do we solve it?
We offer a unique perspective on the behavior of IPv6 scanners
by conducting controlled experiments leveraging a large and 
unused \subnet{56} IPv6 subnet. We selectively make parts of the
subnet visible to scanners by hosting applications
that make direct or indirect contact with IPv6-capable servers
on the Internet. By careful experiment design, we mitigate the 
effects of hidden variables on scans sent to our \subnet{56} subnet
and establish causal relationships between IPv6 host activity types 
and the scanner attention they evoke.
% What do we find?
We show that IPv6 host activities \eg Web browsing, membership in the
NTP pool and Tor network, cause scanners to send a magnitude higher
number of unsolicited IP scans and reverse DNS queries to our subnet than
before. DNS scanners focus their scans in narrow regions of the
address space where our applications are hosted whereas IP scanners
broadly scan the entire subnet. Even after the host activity from our
subnet subsides, we observe persistent residual scanning to portions
of the address space that previously hosted applications.

% \sloppy

\section{Introduction}\label{sec:introduction}

% Set the stage
Scanning the IP address space has exposed security vulnerabilities,
enabling researchers and practitioners to develop effective defenses.
Tools for scanning the IP address space grapple with the fundamental
challenge of efficient scanner target discovery --- a task made more 
challenging by the increasing adoption of IPv6 on the Internet. The IPv6
address space consists of $2^{128}$ possible addresses, rendering brute-force
generation of scanning targets infeasible. Recent work has developed 
tools~\cite{tumi8zma27:online,IPv6Tool84:online} to make Internet-scale 
IPv6 scanning practical by analyzing IPv6 address assignment patterns
~\cite{RFC7707, padmanabhan2020dynamips,plonka2015temporal} 
and developing efficient scanner target generation algorithms
~\cite{RFC7707, padmanabhan2020dynamips,plonka2015temporal}.

% What is the problem/gap?
Despite the recent work on effectively scanning the IPv6 address space, 
little is known about the scanning strategies deployed in the wild. 
We address this gap by analyzing IPv6 scanning from the perspective of
IPv6 hosts on the Internet. Our goal is to reveal target generation
strategies of IPv6 scanners and inform address assignment policies
to mitigate the impact of Internet-scale IPv6 scanners. Previous
work with similar goals performed observational studies using passive 
measurements of unsolicited traffic. While observational studies
provide useful insights they (1) do not identify address
discovery strategies leveraged by scanners and (2) may not be representative of
the real scanning activity observed by actively in-use IPv6 networks
~\cite{ford2006initial, huston2010background, 
czyz2013understanding, fukuda2018knocks} (\Cref{sec:background}). In contrast, 
we take an active approach by conducting \emph{controlled experiments} to evaluate
the impact of IPv6 host activity on scanner behavior.
%  How do we address this gap?
We begin the study by acquiring a previously-unused \subnet{56} IPv6 subnet
owned by a university. This address space did not originate any traffic
prior to the start of the study, allowing us to conduct clean-slate controlled
experiments that make parts of the address space visible on the Internet
for the first time during our study. A subgroup of \emph{treatment} subnets
in our address space host applications that make direct or indirect contact
with potential IPv6 scanners. We measure the \emph{effect} of the treatment 
by capturing unsolicited IPv6 scanning activity received by our address space. 
By comparing the impact on scanning activity between the treatment and \emph{control}
subnets, we establish causal relationships between types of host activities
and increased scanner attention (\Cref{sec:prevalence}).

% Why is this hard?
Accurately associating a measurable increase in scanner attention to 
specific IPv6 host activities is challenging. 
First, due to our large yet limited IPv6 address space for experimentation,
discerning the effect of one host activity (Web browsing) from another (Tor relay)
on increased scanner attention is hard as the observable scanning activity
can result from a combination of host activities. Second, scanning activity
can often persist after the experimental host activity has subsided, confounding
the measured effects of subsequent experiments. Finally, Internet scanners
can coincidentally scan our treatment subnets, endangering false conclusions
about the effect of host activity on scanner attention. We carefully design our 
controlled experiments to mitigate the effects of these hidden variables
to improve the accuracy of our conclusions (\Cref{sec:discovery}).
Our study spans over a year from the time we acquired the \subnet{56} subnet 
and began the controlled experiments. We analyze the
IPv6 scans and reverse DNS queries destined for our \subnet{56} 
address space to make the following key findings:

% What are our contributions?
\para{Host activity has a sizable impact on scanner attention.} 
While our subnet attracted moderate background radiation scans
before we began our controlled experiments to simulate host activity,
the host activity evoked sizable scanner attention --- with IP and DNS scanning
increasing to 225$\times$ and 1.6$\times$ pre-experiment rates, respectively.
On conclusion of our experiments, the rates continued to stay high at
426$\times$ and 3.7$\times$, respectively.
Moreover, host activity that makes direct  contact with IPv6 capable servers on
the Internet (\eg Web browsing,  querying open DNS resolvers) evokes
{50$\times$} and 13$\times$ more activity during and after experimentation, in
comparison with host activities which make indirect contact with potential
scanners (\Cref{sec:prevalence}).

\para{DNS scanners have a narrow attention space.}
Reverse-DNS scanners focus their attention on scanning the
treatment subnets \ie subnets hosting the IPv6 application
and send very few scans to subnets outside the treatment subnet. 
In contrast, IP scanners have a broader attention span, often scanning
both inside and outside the treatment subnets (\Cref{sec:discovery}).

\para{Residual scanning after host activity subsides.}
DNS scanning activity persists long after the host activity that evoked the 
scans has subsided. In specific, we observe high volume DNS scans 
for nearly six weeks in the treatment subnets used for browsing the Web
after the Web browsing activity has ended. The scans re-start
after a break of 2-3 weeks (\Cref{sec:discovery}).

\para{Random and low-byte scanning are dominant strategies.}
We fingerprint the IPv6 scanners that sent unsolicited traffic
to our subnet during the study and find that IPv6 scanners have one 
of two main scanning strategies: they either have equal interest 
in the entire address space (random scanners) or focus more on 
low-byte addresses (\Cref{sec:behavior}).

Finally, we discuss the implications of our findings for network operators (\Cref{sec:security}) and
place our contributions in the context of previous work (\Cref{sec:related}).

\vspace{-1em}
\section{Background}\label{sec:background}
In this section, we provide a high-level overview of IPv6 addressing
and IPv6 scanning strategies.% --- including address discovery and address
%generation strategies.
%   
% \subsection{IPv6 Addressing}\label{sec:background:addressing}

\para{IPv6 addressing.}
An IPv6 address consists of ${128}$ bits which may be broken down into three
parts --- an $m$ bit routing prefix, an $n$ bit subnet ID, and a $k$ (=
128-$m$-$n$) bit interface ID (IID). The routing prefix and subnet ID are used
to route traffic to a local network where hosts are identified by unique IIDs
\cite{RFC4291}. Network operators face two questions while 
allocating IPv6 addresses to networks and hosts (1) how
many bits of an IPv6 address should be allocated to host IIDs (\ie the value of
$k$) and (2) how should IIDs be allocated to hosts in a subnet.

\parait{Determining IID lengths.}
It is considered the best practice for network operators to leave 64 bits for
IIDs according the RFC4291 \cite{RFC4291}. There are many compelling reasons
for this practice. First, a large number of existing IPv6 configuration options
and RFC recommendations assume a 64-bit IID. Therefore, not following this
recommendation may result in operational failure when using IPv6-specific
features and technologies. For example, as pointed out in RFC 4291, a 64-bit
IID is required to use privacy-enhancing IPv6 features which allow for
cryptographically generated addresses (RFC 3972 \cite{RFC3972}), or to use
IPv4-to-IPv6 transition protocols such as 6to4 (RFC 3056\cite{RFC3056}), or to
leverage neighbor discovery protocols implemented in accordance with RFC 4861
\cite{RFC4861}. 

\parait{Assigning IIDs.}
IIDs may be allocated to hosts through (1) manual configuration, (2)
stateless address auto configuration (SLAAC defined in RFC4862) \cite{RFC4862},
or (3) the DHCPv6 IP leasing protocol (defined in RFC 8415 \cite{RFC8415,
richardson2018internet}).
Each of these three methods may result in host addresses having different IID
characteristics. For example, network operators using manual configurations may
assign host IPs in a predictable manner --- \eg using sequential addressing
which results in the lower bytes of the IID being populated with all leading
bytes set to 0 (as per RFC7707 \cite{RFC7707}) or assign host IP addresses
based on their 48-bit MAC addresses (the modified EUI-64 protocol defined in
RFC4291 \cite{RFC4291}). On the other hand, operators using SLAAC or other
approaches (\eg from RFC 7943 \cite{RFC7943}, RFC 7217 \cite{RFC7217},
and RFC 3972 \cite{RFC3972}) generate pseudo-random IIDs. 

\parait{Our address allocation approach.}
Given the importance of a 64-bit IID length, it is reasonable to assume that
scanners assume 64-bit IIDs in their scanning targets. Therefore, in all our
subsequent experiments detailed in \Cref{sec:prevalence} and
\Cref{sec:discovery}, we assign 64-bit IIDs to all our hosts and subnets.
Further, we use both IID generation approaches (lower-byte and pseudo-random
addresses generation) to generate IPs for hosts whose addresses are leaked to
scanners. This allows us to study the impact of address allocation methods on
the subsequent address generation strategies leveraged by IPv6 scanners --- \eg
does the discovery of a host with a pseudo-random (or, lower-byte) IID result
in scanners only sending probes to other addresses with a pseudo-random (or,
lower-byte) IIDs.

\para{IPv6 address representation.} A common method for representing IPv6
addresses uses 32 \emph{nybbles}, each denoted in hexadecimal and
representing four contiguous bits. Every four nybbles (two bytes) are
separated by a `:' and any leading zeros in these two byte sections may be
dropped. 
%
% Therefore, {\tt 2601::dead:1}, {\tt 2601::dead:0001}, and {\tt
% 2601:0000:0000:0000:0000:0000:dead:0001} all represent the same address. We
% refer to the last expression as the \emph{exploded} representation of the
% address.

\parait{Representation in DNS reverse zones.} DNS reverse zones map addresses
to domains --- \ie the opposite of what DNS zones do. Sending a DNS PTR query
to the corresponding IPv6 reverse zone returns the domains hosted with the
corresponding IP address. IPv6 PTR records are organized under {\tt ip6.arpa}
in 32 levels where each nybble of an exploded IPv6 address is a level.
Therefore, the PTR record associated with the address {\tt 2601::dead:1} is
available at {\tt 1.0.0.0.d.e.a.d.<20 repeating .0s>.1.0.6.2.ip6.arpa}.

\para{Scanning strategies: address discovery.} Unlike IPv4, scanning the entire
IPv6 address space for live addresses is infeasible. Instead, scanners focus
their attention on regions of the address space near addresses where some
activity has been observed. Prior work has used techniques like monitoring 
public lists of IPv6 addresses (\eg TLD zone
files which list the IPv6 addresses of domains) or hosting public services to
receive contact from previously unseen addresses (\eg hosting a web
service). % Each of these sources have been leveraged in
% prior work (described in \Cref{sec:related}) to discover IPv6 scanning
% targets. 

\parait{Our address leaking strategies.}
In our controlled experiments described in \Cref{sec:prevalence} and
\Cref{sec:discovery}, we leak the `liveness' of specific regions of an IPv6
network by the following means: (1) direct contact with IPv6 capable web
services, (2) sending DNS queries to public DNS resolvers, (3)
participation in the NTP pool protocol \cite{poolntpo1:online} where IP
addresses of participants can be enumerated \cite{WhatisNT71:online}, (4)
participation in the NTP public server protocol \cite{WebHomeS5:online} where
our addresses are distributed via a public listing of NTP servers,
(5) participation in the Tor network as a middle-relay where our addresses are
distributed via the Tor consensus \cite{Sources34:online}, and (6)
registering domains with specific addresses so their presence is
known via the TLD zone files. 

\para{Scanning strategies: probing for liveness} Once scanners have a specific
region of the address space to focus on, they may use a different strategy to probe
for active hosts. 

\parait{Traditional IP scanning.} A common approach is to solicit responses
from live hosts by sending probes (packets
associated with a common Internet protocol such as ICMPv6) to a set of
candidate IP addresses using tools such as \emph{zmapv6}
\cite{GitHubtu35:online} and \emph{nmap} \cite{IPv6Scan40:online}. These
candidates are generated by observing patterns in the already identified live
addresses (\eg known live addresses are sequentially allocated) by
off-the-shelf tools like \emph{ipv666} \cite{GitHubla62:online} and
\emph{the IPv6 toolkit} \cite{IPv6Tool92:online}.

\parait{NXDOMAIN scanning.} An emerging and increasingly popular approach for
scanning IPv6 spaces for liveness involves leveraging a feature of the IPv6
reverse-DNS lookup process detailed in RFC8020 \cite{RFC8020}. RFC 8020
\cite{RFC8020} states that reverse-DNS lookups within subnets that contain no
domains should receive an {\tt NXDOMAIN} response code --- \eg if the prefix
{\tt 2601::dead:1/80} contains no domains, reverse-DNS queries for any
more-specific prefixes should return an {\tt NXDOMAIN} response code. This
allows scanners to make significant reductions the address search space. 

\parait{Our data logging approach.} In our experiments, we are interested in
identifying and detailing the behavior of traditional IP scanners and NXDOMAIN
scanners. Therefore, we set up our infrastructure to capture all packets and
DNS PTR lookups for our address space.

\section{Prevalence of IPv6 Scans}\label{sec:prevalence}
In this section, we answer the question: \emph{How prevalent is
IPv6 scanning in the wild?} 
% We describe our methodology for measuring
% IPv6 scanning activity (\Cref{sec:prevalence:methodology}),
% characterize the scans (\Cref{sec:prevalence:results}) and the scanners
% (\Cref{sec:prevalence:characteristics}). 
% This serves as the basis of our analysis in \Cref{sec:discovery} and \Cref{sec:behavior}.

\subsection{Data collection methodology}\label{sec:prevalence:methodology}
\para{Overview.} 
Our study is based on scanner behavior observed on a previously unused and
unannounced \subnet{56} IPv6 address space (\Cref{sec:prevalence:methodology:chars}). 
We conduct controlled experiments by creating
a \emph{treatment group} of \subnet{64} subnets which contain publicly
advertised and visible services and a \emph{control group} of subnets with no
services (\Cref{sec:prevalence:methodology:attention}). 
To log scanning behavior, we use a logging infrastructure
that captures all packets and DNS queries sent to our \subnet{56} IPv6 address
space (\Cref{sec:prevalence:methodology:logging}). 
%
% We analyze the packet captures and DNS logs to quantify the
% prevalence of IPv6 scanning activity on the Internet. We compare the scanning
% activity observed by the treatment group with that of the control group to
% understand (1) the impact of deploying specific public-facing services on IPv6
% scanning activity within and outside an IP address space and (2) the methods
% used by IPv6 scanners to discover scanning targets.

\subsubsection{Characteristics of our IPv6 address space}
\label{sec:prevalence:methodology:chars}

\begin{figure}[t]
  \centering
  \includegraphics[width=0.35\textwidth]{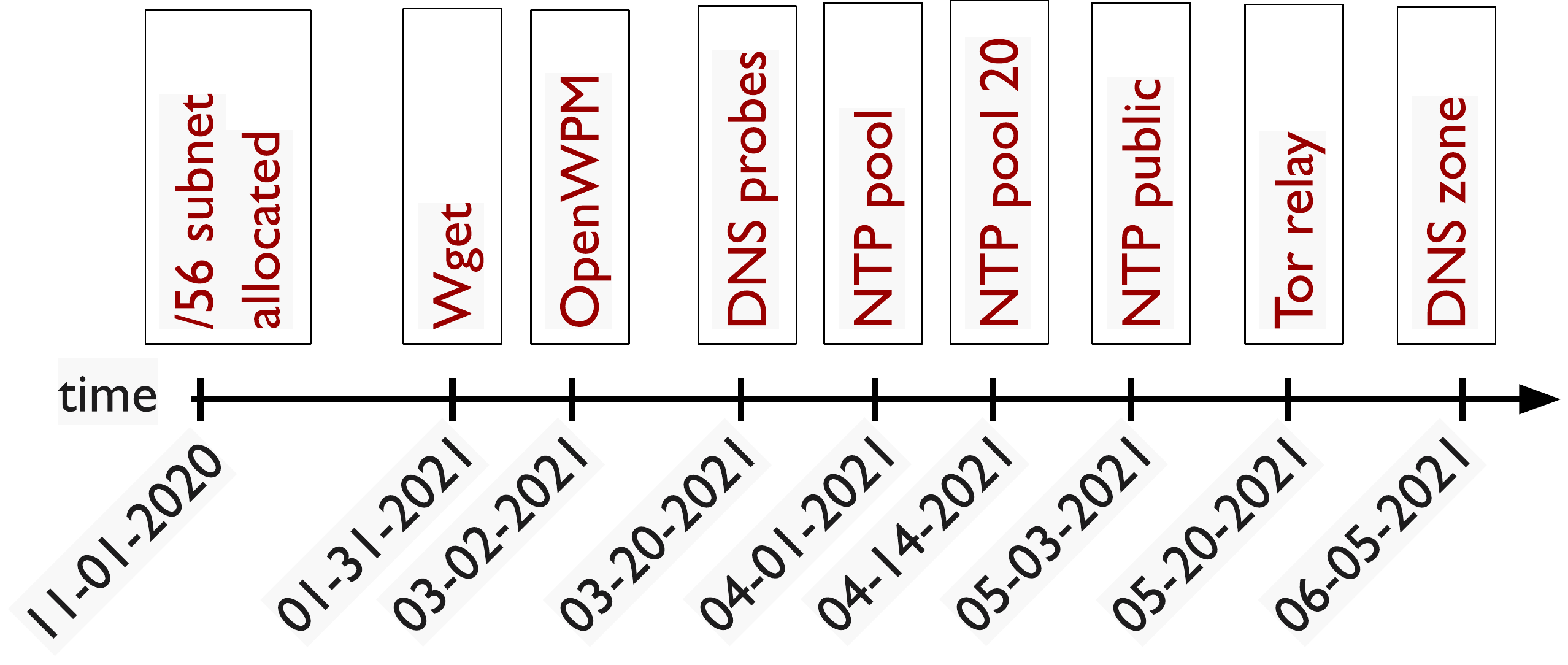}
  \caption{Timeline of our experiments that simulate host activity from a \subnet{56}
IPv6 subnet we own.}
  \label{fig:timeline}
\end{figure}

Our \subnet{56} subnet is a part of an autonomous system (AS)'s 
\subnet{48} allocation. Since the \subnet{56} subnet was previously unused and unannounced by
BGP, it should not receive any legitimate traffic~\cite{caidatelescope}. 
We were granted access to the \subnet{56} in {\it 11/2020}
after which the parent AS announced it to the Internet via BGP and we setup
data logging infrastructure (\Cref{sec:prevalence:methodology:logging}). We left
the address space idle for the next three months to benchmark base levels of
Internet background radiation received by our subnet. In {\it 03/2021}, we
began a series of controlled experiments to understand IPv6 scanner behavior. 
Figure~\ref{fig:timeline} summarizes the timeline of our experiments.

\para{Separating our \subnet{56} into treatment and control groups.}
We run six different controlled experiments on our address space ---
each requiring four unique end-host IPs. We first sub-divided the \subnet{56}
address space into four \subnet{58} address spaces. 
Then, due of the importance of using 64-bit IIDs for 
each host ({\em Cf.} \Cref{sec:background}),
we allocated each of our 24 end-hosts to a unique \subnet{64} subnet and
assigned them addresses from this subnet. Therefore, each of our six
experiments were conducted on four unique end-hosts contained in four unique
\subnet{64} subnets. For two of the four end-hosts associated with each
experiment, we allocated a pseudo-random IID. The remaining two received
lower-byte IIDs. \emph{Thus, we conduct each experiment on two end-hosts
which reflect the two most common forms of address assignment in IPv6
networks.}
The 24 \subnet{64} subnets associated with our experiments are
our \emph{treatment subnets} and every other subnet is a \emph{control subnet}.
All treatment subnets were randomly chosen from the same \subnet{58} subnet
such that no two were adjacent to each other. This allows for proper
analysis of treatment effects (\Cref{sec:discovery}) and facilitates retries
on the remaining three \subnet{58} subnets if an experiment
had to be repeated due to failures. Fortunately, the latter was not
required.

\subsubsection{Attracting scanner attention}
\label{sec:prevalence:methodology:attention}

Following the initial three month period of inactivity from 11/2020 to 03/2021,
we simulated host activity from our treatment subnets by launching {\em
services} (\ie experiments\footnote{In the remainder of this paper, experiments
and services are used interchangeably.}) that mimic specific types of host
behaviors on the Internet. 
We ran one service at a time on the four corresponding treatment subnets for
\emph{at least} 2 weeks to measure the impact on scanning behavior caused by
the \emph{specific host activity} that is mimicked by the service. More details
on the methods for measuring service effects are in \Cref{sec:discovery}. 
%
% Between any pair of experiments, there is a cool-off period of at least 2 weeks. 
% The cool-off period serves two purposes. It captures any delayed scans 
% resulting from the previous experiment and separates the scans resulting 
% from the two experiments at either end of the cool-off period.
% While a longer cool-off period can benefit both these goals, 
% it would increase the duration of our study by a magnitude.
%  
% \Cref{fig:timeline} shows the timeline of our experiments. We ran four
% simultaneous instances of each experiment with each instance allotted  to
% a different IPv6 address. We do not reuse IPv6 addresses after they have been
% used in an experiment. For each experiment, we ran 2 instances on lower-byte
% interface ID (IID) addresses and the other 2 instances on random IID addresses.
% A lower-byte IID address has all 0s in the interface ID part of the IPv6
% address except for the last byte \eg \textit{2001:d00f:1:2::23} is an
% abbreviated version of \textit{2001:d00f:1:2:0000:0000:0000:0023}. A random IID
% address does not have a definite pattern in the interface ID part of the
% address \eg \textit{2001:d00f:1:2:83c4:4c0:bb21:8766}. We placed the
% 4 addresses used for a single experiment in distinct and randomly chosen
% \subnet{64} subnets. These four subnets have a gap of at least one \subnet{64}
% prefix between them (\rsnote{Figure 2}). 

\para{Experiments (services) deployed.}
The goal of our experiments is to identify the effect of IPv6 host activity 
on scanning behaviors. We achieve this goal by simulating six types of host
activity from the 24 \subnet{64} treatment subnets. 
Each experiment \emph{uses a different method to leak the liveness
of the treatment \subnet{64} subnets} to scanners. These methods are based on
findings from prior work which highlight the sources of IPv6 addresses
leveraged for efficient IPv6 scanning (\Cref{sec:related}).
We leak our addresses using a combination of \emph{direct} and
\emph{indirect} scanner contact approaches. Direct contact approaches send
packets directly to IPv6 addresses with the expectation of receiving scanner
attention in return. In comparison, indirect contact approaches enlist our
services in public lists that may be monitored by scanners seeking to discover
new IPv6 addresses. 
Our six experiment deployments are described below.

\parait{Experiment 1: Web crawls to popular websites.} 
With this direct contact experiment, we mimic web browsing from
a standard home network where users make connections to web
servers that scanners may be operating or tapped into for sources of IPv6
addresses \cite{richter2019scanning}.
We first identified all IPv6-capable websites in the Alexa Top 10K websites
obtained in 02/2021 by collecting their {\tt AAAA} DNS records and checking
them for validity. In total, we found 2.6K IPv6-capable websites which were the
subject of our crawls.
Following the recommendations of Ahmad \etal \cite{ahmad-www2020}, 
we conducted crawls using a simple CLI crawler which did not load
third-party or dynamic content ({\tt wget}) and a full-fledged browser using
OpenWPM \cite{openwpm}.
Our {\tt wget} crawls did not load third-party content and therefore only
established direct connections with the web servers of each website while the
OpenWPM connections used Firefox to also load dynamic content and make
connections with all third-party web servers associated with a website.
Therefore, each crawl leaked the same four host addresses to a different (but
overlapping) set of web servers. Each crawl was conducted 2 weeks apart.

\parait{Experiment 2: Querying DNS open resolvers.} 
In this direct contact experiment, we leak our treatment subnet
liveness to open IPv6-capable DNS resolvers. Since no such list of
resolvers exists for IPv6, we used the approach of Hendriks \etal
\cite{Hendriks-pam2017} to identify IPv6-capable resolvers from IPv4 open
resolver lists. 
In total, we obtained 9K IPv6-capable open resolvers and queried each of them
for the {\tt AAAA} record of {\em www.google.com}. These queries were repeated
every day for a two week period. Therefore, this experiment leaked the liveness
of four treatment subnets to over 9K IPv6 open resolvers.

\parait{Experiment 3: NTP pool servers.} 
In this indirect contact experiment, we hosted four instances of NTP pool
servers in four treatment subnets. 
%
% We found that despite binding the NTP pool servers to different host addresses
% on the same Linux machine, the NTP implementation used a single egress IP
% address for all four instances. This defeats the purpose of exposing four
% different parts of the subnet through the experiment. 
% 
To ensure that each NTP pool instance used a different egress IP address, we
created four network namespaces on our Linux machine to isolate the NTP servers
we hosted. Network namespaces ensure separate ports and IP addresses are assigned
to them, allowing each of our NTP pool servers to use a different IP address
associated with one of the four treatment subnets allocated to this experiment. 
Each server was initially configured with the NTP default parameters which set
rate limits on our responses to liveness probes from other NTP servers for the
first two weeks. These rate limits prevented it from achieving the maximum pool
score of 20 during this period. During this time, the server was usable by
clients (and therefore discoverable by scanners) but not recommended due to
a low pool score. In \Cref{sec:discovery} we refer to this part of the
experiment as `NTP$_{pool}$'.
We removed the rate-limit after two weeks and consequently our servers
immediately achieved the maximum pool score of 20 and was recommended for
client use. We carried this phase of the experiment for another two weeks and
refer to it as `NTP$_{pool-20}$'.
Note that NTP pool servers are not publicly listed on a website, but are
possible to enumerate \cite{WhatisNT71:online}. Therefore, this indirect
experiment leaked the liveness of its four treatment subnets to scanners that
enumerate NTP pool server lists for scanning destinations.
% 
% We hit another complication when the rate 
% limits on \rsnote{what?} prevented our NTP pool servers from responding 
% to liveness probes sent by other NTP servers. Due to this, the NTP pool 
% score of our servers was below the threshold required for the servers to 
% be used by NTP clients. We identified and fixed this issue in roughly 
% 14 days. After resolving this issue, our servers reached the NTP pool 
% score of 20, making them usable by NTP clients. 
% which the score of 4 of pool
% servers was consistently 20. Not publicly listed on a globally visible.
% Limited visibility. More effort to enumerate all non-public servers

\parait{Experiment 4: NTP public servers.} 
While our NTP pool servers were used by clients for synchronizing time, they
were not publicly listed and require additional effort to enumerate. In this
indirect contact experiment, we launched four NTP {\em Stratum 2} public
servers that remained active for two weeks.
Unlike pool servers, these are published on an archived list making them more
visible to scanners \cite{poolntpo47:online}. 
This indirect contact experiment leaked the liveness of its four
treatment subnets to scanners monitoring NTP server lists.

\parait{Experiment 5: Tor relays.} 
For this indirect contact experiment, we launched four Tor~\cite{torpaper}
middle relays with unrestricted bandwidth in our subnet. These relays remained
operational for a two week period.
Since our main purpose was to enlist on the public Tor consensus, we
chose middle relays as opposed to entry or exit relays. We made this
decision because entry relays receive information about the clients connecting to
Tor and exit relays receive information about the destinations of Tor traffic.
We did not find such information appropriate to gather and analyze. For the
additional safety of Tor users, we discarded any non-scanner traffic (defined
in \Cref{sec:prevalence:methodology:logging}) destined for the deployed relays.
Therefore, this indirect contact experiment leaked the liveness of the four
treatment subnets to any IPv6 scanners monitoring the Tor consensus.

\parait{Experiment 6: DNS zone files.} Finally, we registered
four domains, two each with a \texttt{.com} and \texttt{.net} TLD. The AAAA
records of all four domains pointed to addresses from four of our treatment
subnets. 
Registering these domains with the {\tt com} and {\tt net} TLDs results
in them getting added to the largest TLD zone files. Since prior work has
leveraged these lists to identify web services with IPv6 addresses, 
we expect to make indirect contact with scanners monitoring these 
zone files.

\para{Limitations of our experimental setup.}
We selected the above-mentioned services for two reasons: (1) 
they provide a range of common host activities that expose 
their IP addresses to potential scanners and (2) based on their
use in IPv6 target generation and address extraction 
in prior work~\cite{RFC7707, gasser2018clusters, Targeted49:online}. 
These services allow us to "leak`` our address space to scanners 
and achieve two goals. First, they help us identify how 
scanning behaviors can affect commonly used services and second, 
they provide a method for rigorously measuring the exposure of a 
specific service to scanners. 
We note that our choice of deployed services is not comprehensive --- 
i.e., it is possible that other services attract different
scanning patterns. 

\begin{figure}[t]
    \centering
    \begin{subfigure}[b]{0.49\textwidth}
        \includegraphics[width=\textwidth]{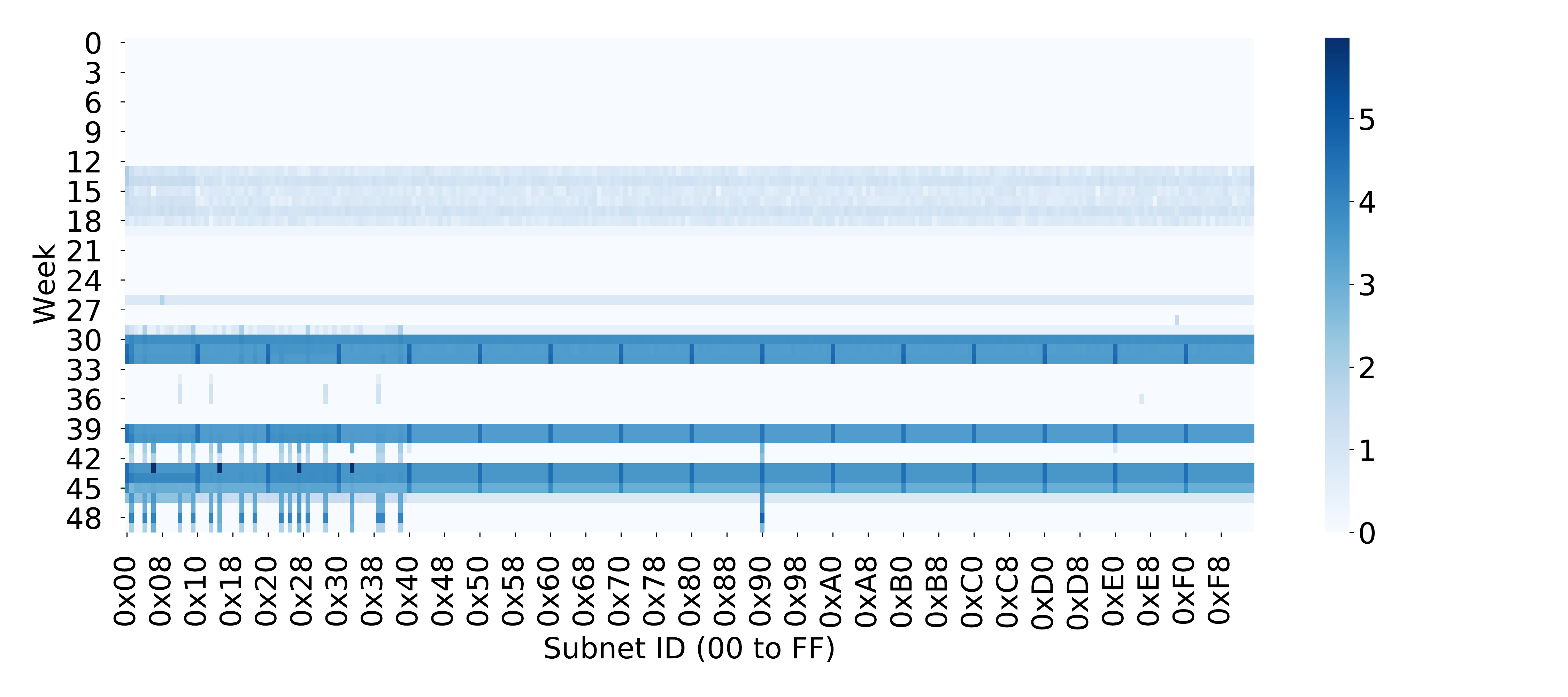}\\
        % \caption{IP scanning activity}
        \label{fig:prevalence:results:ipheatmap}
    \end{subfigure}
    %\vspace{-2em}
    \begin{subfigure}[b]{0.49\textwidth}
            \includegraphics[trim=0cm 0cm 0cm 0cm,width=\textwidth]{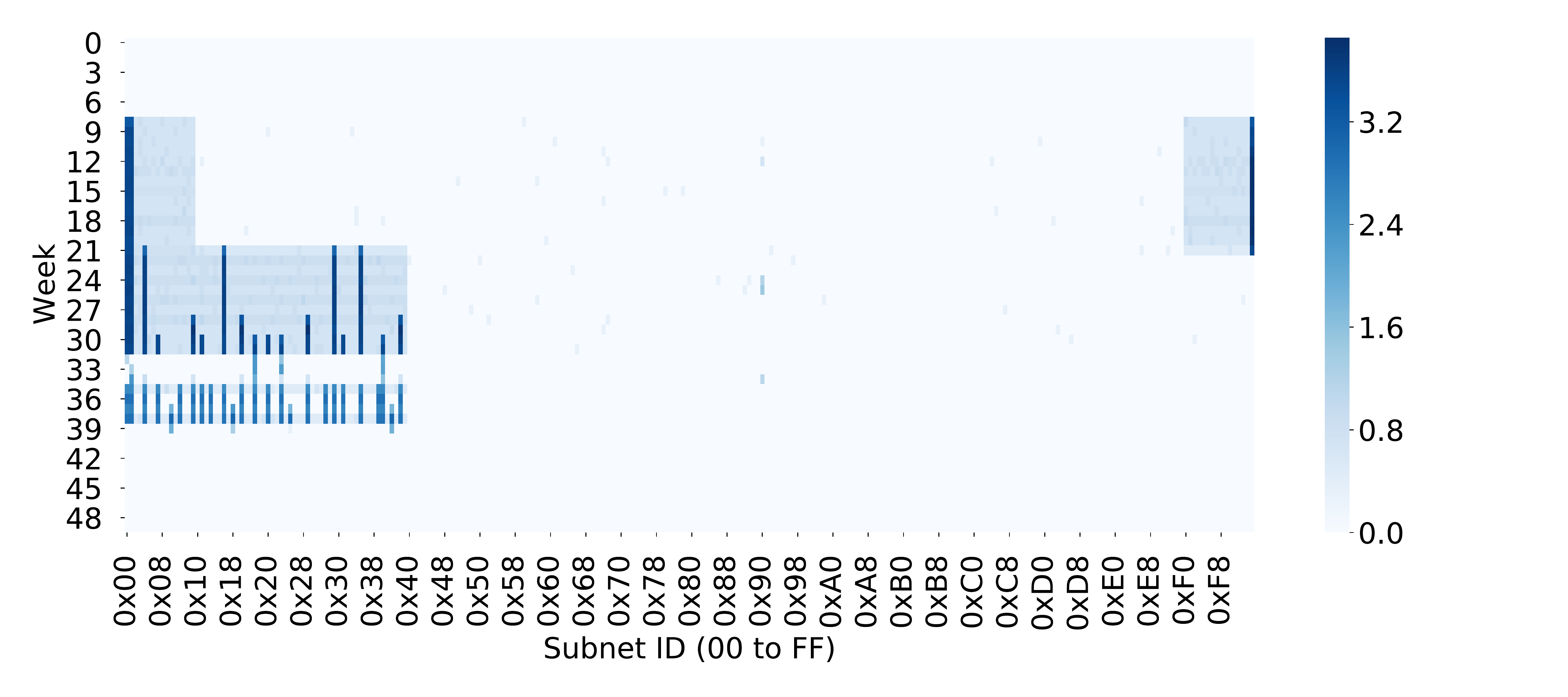}\\
            % \caption{NXDOMAIN scanning activity}
            \label{fig:prevalence:results:dnsheatmap}
    \end{subfigure}
    %\vspace{-1em}
    \caption{Scanning activity observed by our 256 \subnet{64} subnets during
    each of the 48 weeks in our study. IP scans shown in the top and DNS 
    scans in the bottom figure. Values are $\log_{10}$-scaled.
    \label{fig:prevalence:results:heatmaps}}
\end{figure}

\subsubsection{Measuring scanning activity}
\label{sec:prevalence:methodology:logging}
\para{What is a scanner?}
Our experiments solicit legitimate traffic amongst traffic from
scanners (\eg NTP clients may send legitimate queries to our NTP pool servers).
To differentiate between traffic from legitimate sources and scanners,
we analyze the sources of all traffic and label scanners as any
sources that send packets to an IP address not associated with any of our
experiments. Further, we take a conservative approach and: (1) count all
sources emerging from the same \subnet{64} subnet as a single scanner --- we do
this because our data shows several scanners using a distributed infrastructure
for scanning (identified by sequential source addresses and wide area scans)
and (2) also discount traffic from sources we previously communicated with as
part of our experiments. While this results in removing scanners
using the same IP address as the service they run (\eg public DNS resolver), it prevents
accidental over counting (\eg of NTP clients that
need to perform reverse-DNS lookups as part of the protocol
\cite{poolntpo47:online}).

\para{Data logging.} 
%It was important to gather scanning traffic associated with traditional IP
%scanning as well as NXDOMAIN scanning. 
To capture IP scanning, we captured
PCAPs of all traffic originating from and destined to any address in our
\subnet{56} allocation. 
To capture NXDOMAIN scanning, our AS ran an authoritative DNS server which
served all domains assigned to our IP address allocation with {\tt TTL = 1}.
This low TTL ensured that PTR queries \cite{rfc1035} were not cached and
continue to reach our authoritative server. We were provided with daily dumps
of logs from the server which contained the timestamps, sources, and
destinations of queries received.

\begin{table}[t]
  \centering
  \small
  \scalebox{\tabularscale} {
  \begin{tabular}{lccc}%p{.9in}p{.9in}}
    \toprule
      &   & {\bf DNS logs}  & {\bf PCAPs} \\
    \midrule
    \# Scanner probes & Before & 200,335  & 13,044 \\
                      & After  & 499,638  & 14,564,017 \\
                      & Total  & 699,973  & 14,577,061 \\
    \midrule
    \# Scanners       & Before & 20       & 5 \\
                      & After  & 89       & 1065 \\
                      & Total  & 96       & 1068 \\
    \bottomrule
  \end{tabular}
  }
  \caption{Activity observed by our address space before and after experiments
  began from PCAPs and DNS logs.}
  \label{tab:prevalence:summary}
\end{table}

\subsection{Prevalence of scanning}  
\label{sec:prevalence:results}

During the 48 week period of our study, our subnet received 14.6M packets and
699K DNS queries from 1068 traditional IP scanners and 96 NXDOMAIN
scanners, respectively. We provide a summary of these scans in
\Cref{tab:prevalence:summary}. 
Table~\ref{tab:prevalence:summary} shows the number of IP and NXDOMAIN
scans observed by our subnet before and after our experiments began, split by
the treatment and control subnets. We see that, both control and treatment
subnets appear to receive only small amounts of scanning activity per day
before our experiments began in February 2021 from NXDOMAIN and IP scanners.
Following the start of our controlled experiments, we see large increases
observed in both groups of subnets --- with some days receiving upwards of
100K IP scans and 1K NXDOMAIN scans. The plots are not normalized by 
the number of subnets. 
%
%
%\para{Takeaways.} 
\Cref{fig:prevalence:results:heatmaps} shows the amount of scanning activity
destined for each of our 256 \subnet{64} subnets.
First, we see a clear difference
in the way IP scans and NXDOMAIN scans are used. Specifically, NXDOMAIN
scans are narrowly focused on a smaller region of our address
space while IP scans broadly target the entire \subnet{56} region. This
attests to how the NXDOMAIN semantic can be used to discard entire subtrees
 and focus scanning attention on regions with active 
\subnet{64} subnets.
Second, we see that not all subnets receive the same amount of attention from
scanners --- some receive either an increased number of scans or are scanned
for longer durations of time. Third, we see that scanning activity is not
consistent, there are several quiet weeks even after the experiments
begin. In the next sections, we investigate the causes for these 
differences --- \ie what attracts scanners to
specific subnets? (\Cref{sec:discovery}) and how do
scanners find active subnets and what happens once they find one?
(\Cref{sec:behavior}).

\subsection{{Scanner characteristics}}
\label{sec:prevalence:characteristics}

\begin{figure*}[t]
  \centering
  \begin{subfigure}[t]{.325\textwidth}
    \includegraphics[width=\textwidth]{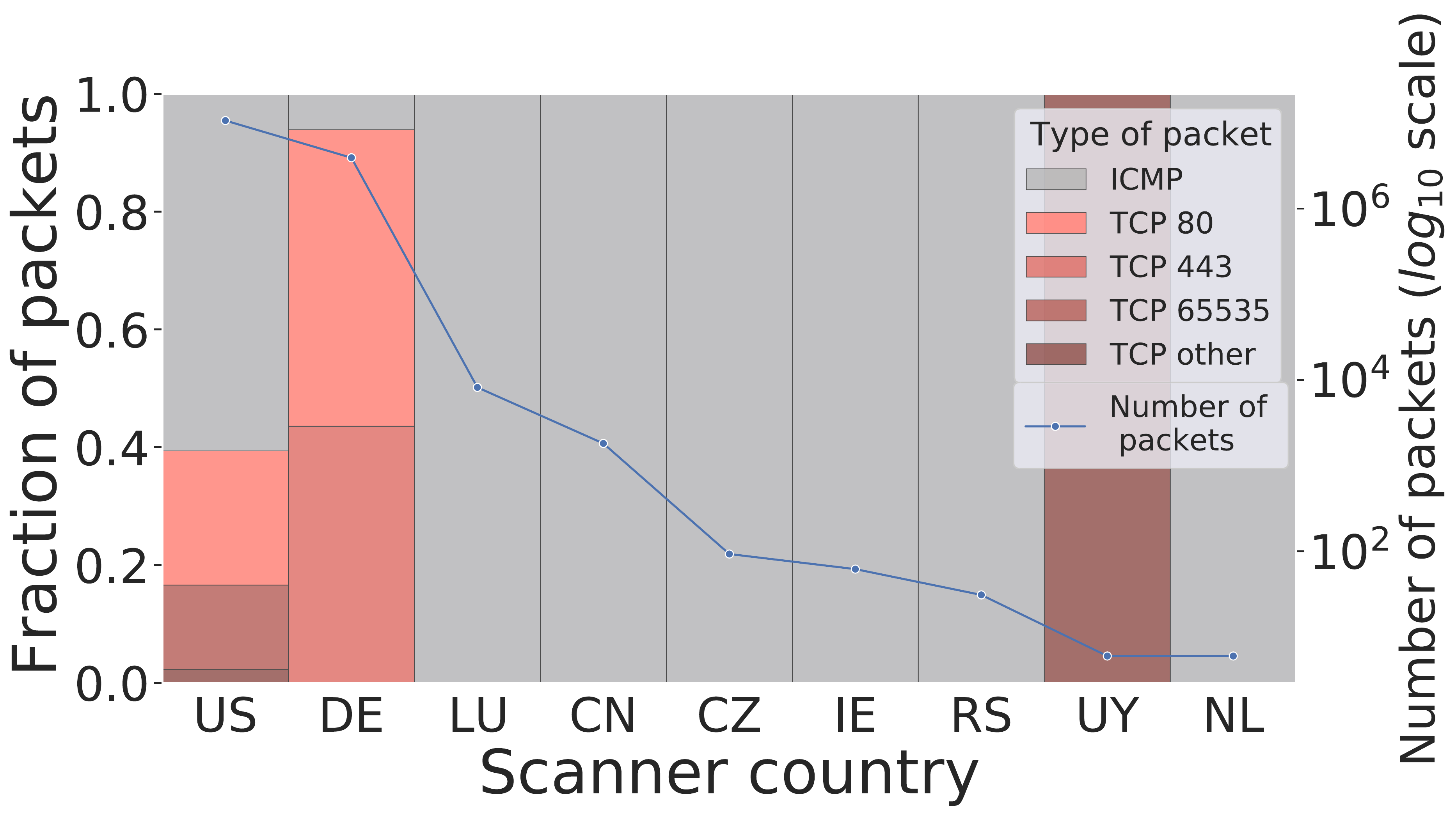}
    \caption{IP scanning by country}
    \label{fig:ip_characterization:country}
  \end{subfigure}
  \begin{subfigure}[t]{.31\textwidth}
    \includegraphics[width=\textwidth]{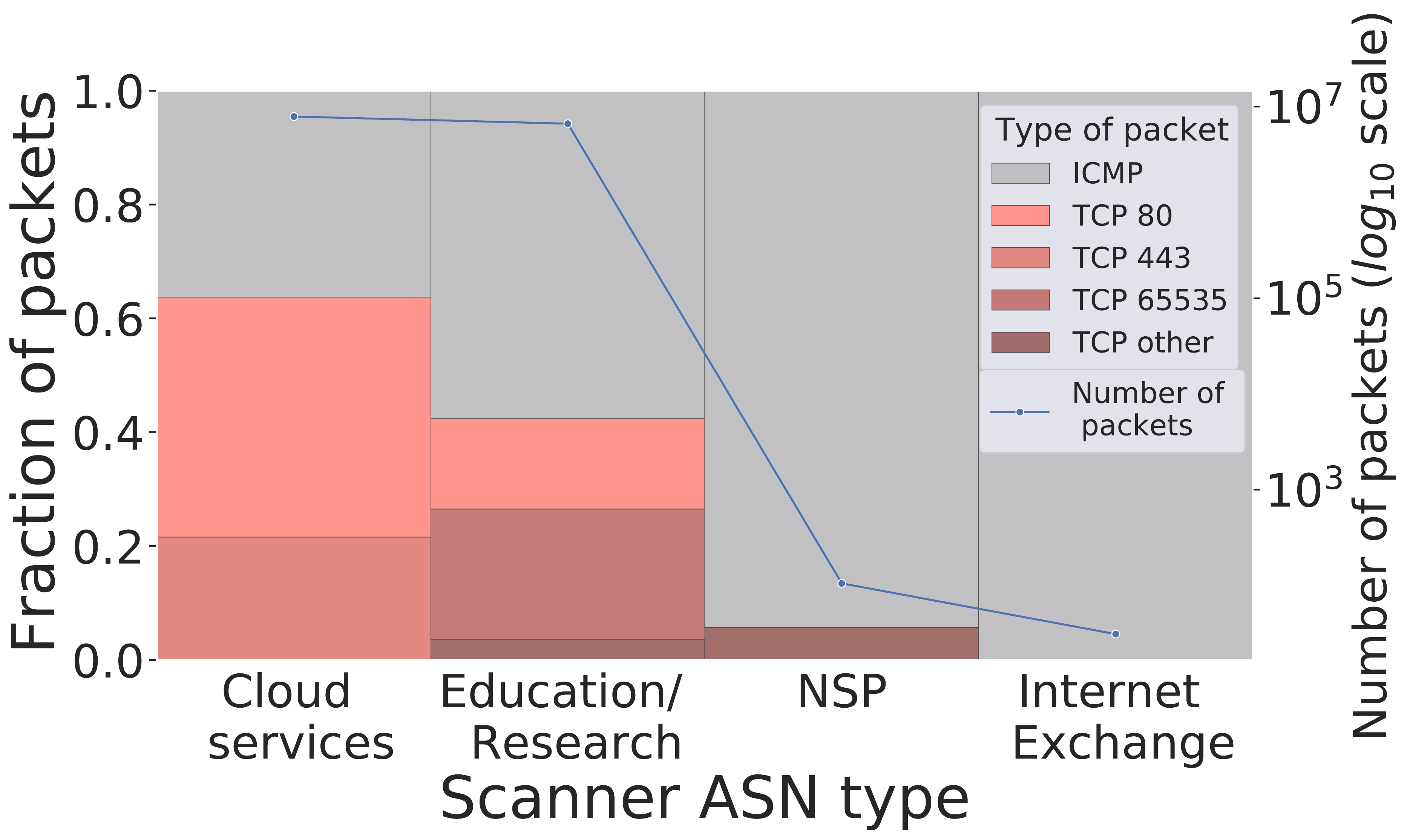}
    \caption{IP scanning by ASN type}
    \label{fig:ip_characterization:asn}
  \end{subfigure}
  \begin{subfigure}[t]{.325\textwidth}
    \includegraphics[width=\textwidth]{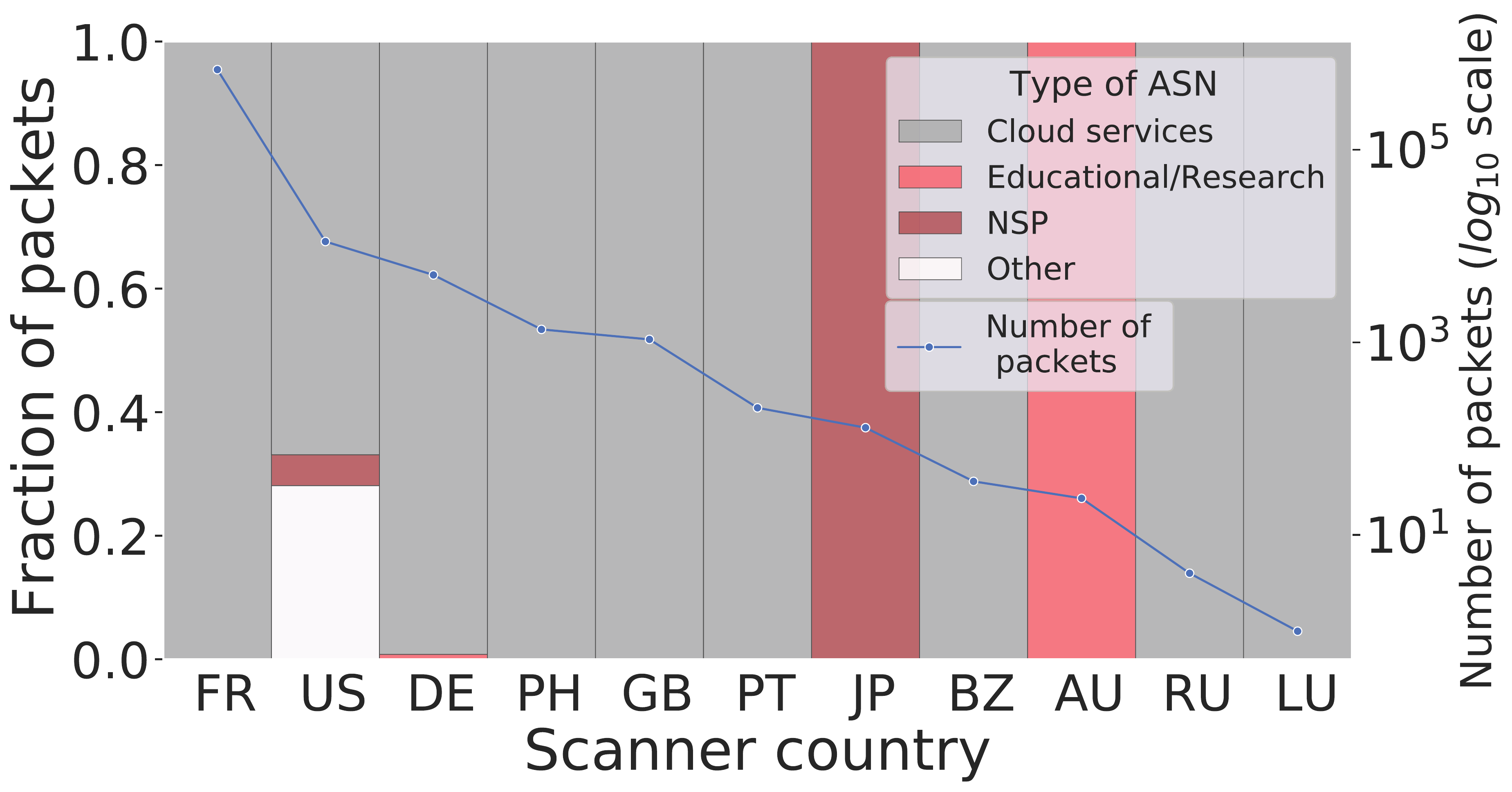}
    \caption{NXDOMAIN Scanning by country}
    \label{fig:dns_characterization}
  \end{subfigure}
  \caption{Breakdown of scanning traffic by country and ASN type.}
\end{figure*}

% \begin{figure}[t]
%     \centering
%     \begin{subfigure}[b]{0.5\textwidth}
%         \centering
%         \includegraphics[width=\textwidth]{plots/resub_plots/scanner_characterization.pdf}
%     \end{subfigure}
%     \hfill
%     \begin{subfigure}[b]{0.5\textwidth}
%         \centering
%         \includegraphics[width=\textwidth]{plots/resub_plots/scanner_characterization_asn.pdf}
%     \end{subfigure}

%     \caption{Breakdown of IP scanning traffic by country and ASN type}
%     \label{fig:ip_characterization}
% \end{figure}

\para{Scanner sources.} We show the distribution of IP scanning
traffic categorized by originating country (\Cref{fig:ip_characterization:country})
and type of AS (\Cref{fig:ip_characterization:asn}). Over 99\% of 
all scanning traffic originated from just 2 countries,
the US and Germany. Similarly, 99\% of all scanning traffic originated 
from ASes belonging to cloud service
providers or education/research institutes. 
We contacted the most active scanner in the education/research category 
of ASes to inquire the purpose of their scans. They responded that 
their goal was to ``measure the state of IPv6 adoption''.
Other ASNs in this category belonged to research institutes including
\textit{China Next Generation Internet} which 
is tasked with with the ``transition from IPv4 to IPv6'' in China 
\cite{IFTFIPv679:online}. Of the scanning traffic that originated 
from cloud service providers, >99\% of traffic originated from 
\textit{Digital Ocean} and \textit{ScaleUp Technologies}.
% \begin{figure}[t]
%     \centering
%     \includegraphics[width=0.5\textwidth]{plots/resub_plots/dns_scanner_characterization.pdf}
%     \caption{Breakdown of NXDOMAIN scanning traffic by country and ASN type}
%     \label{fig:dns_characterization}
% \end{figure}
\Cref{fig:dns_characterization} breaks down NXDOMAIN scanning by 
country and AS type. We observe a similar pattern of two 
countries accounting for 99\% of all NXDOMAIN scanning --- France and the US. 
However, NXDOMAIN scanning originates predominantly only 
from ASes belonging to cloud service providers (> 99\%), with 97\% originating
from the {\it OVH Cloud}. 
ASes belonging to education/research institutes comprise only
a fraction of the total scanning traffic. %~97\% of all NXDOMAIN 
%scanning activity originated from \textit{OVH cloud}.

\para{Scanner traffic characteristics and payload analysis.}
\Cref{fig:ip_characterization:country} and \Cref{fig:ip_characterization:asn} 
also show the breakdown of types of scanning traffic by country and type 
of AS. ICMPv6 (46\%) and TCP SYN (54\%) scans dominate with 
almost all the scans comprising of these two types of traffic. 
%ICMPv6 traffic consists 46\% of all scanning traffic and the 
%other 54\% being TCP traffic. 
Of the TCP traffic, three ports (HTTP, HTTPS, and port 65535) accounted for
97\% of scanner probe targets.   
Of the 14 scanners that did not target these 3 services, 
2 scanners conducted a port scan of the subnets running 
{Tor relays}. Services targeted by these port scans 
included {Telnet, SMTP, FTP}. The remaining scanners 
conducted a wider port scan targeting a total of 1,323 
different ports. All 12 of these scanners originated from the same 
AS and sent very similar scanning traffic in terms of patterns 
in target addresses. These scanners targeted a total of 1,323 
ports including those of popular services like {Telnet, IMAP, PPTP, 
POP3, BGP}.
Of the $14.6$ million scanning packets we captured,
only 4K contained a payload. Packets that did not contain 
any payloads were either TCP SYN packets or ICMPv6 Echo Requests.
Packets containing payloads were either {HTTP} or 
{NTP} packets. We analyzed the {HTTP} packets and found that
one of the scanners had a distinctive {\tt GET} request associated with an Nmap
scan \cite{IPv6Scan83:online}. Further inspection showed that this scanner used
Nmap to scan for potentially open services on 
port 80 like HNAP (Home Network Administration Protocol) with 
well-known exploits \cite{CVE-2021-34829}. NTP packets did 
not seem to contain malicious payloads.

\para{Presence of IP scanners on IP blocklists.}
To understand the intentions of the IP scanners, we examine whether 
they sent unsolicited traffic to other IP addresses on the Internet.
We analyzed if the addresses of scanners targeting our address space
were present on IP blocklists. We used AbuseIPDB \cite{AbuseIPD29:online} 
which maintains reputation of IPv6 addresses. 
Out of the 1,068 scanners observed in our measurements, 12 were
reported for potential abuse on AbuseIPDB. 1,037 of the scanners 
originated from the \subnet{32} IPv6 allocation of the 
AS \textit{Alpha Strike Labs GmbH} which conducts global scans for research.
The 12 scanners that were present on the blocklist contributed 51\% of 
total IP scanning traffic. For 5 of these, we received their scans 
before they were reported on AbuseIPDB. Other scanners 
were reported on AbuseIPDB at most 3 months before they scanned our IPv6 
subnet. All the reports submitted for these scanners mentioned unsolicited 
port scanning.  

\para{Temporal differences in targeted protocols.}
Table~\ref{tab:prevalence:temporal} shows the evolution 
of targets of IPv6 scanning over time. While the techniques 
used in previous work to detect scanning traffic in IPv6 might 
differ from ours, all studies capture traffic that is 
unsolicited/unclaimed. We observe that IP scanning traffic 
has shown similar characteristics over the years with ICMPv6 Ping, 
HTTP (80) and HTTPS(443) being the main targets.

\begin{table}[]
  \centering
    \resizebox{0.4\textwidth}{!}{%
    \begin{tabular}{|llll|}
    \hline
    \textbf{2013 \cite{czyz2013understanding}}                     & \textbf{2018 \cite{fukuda2018knocks}}                    & \textbf{2020 \cite{strowes2020debogonising}}                     & \textbf{2022 (this paper)} \\ \hline
    \multicolumn{1}{|l|}{ICMPv6 Ping} & \multicolumn{1}{l|}{HTTP (80)}   & \multicolumn{1}{l|}{HTTP (80)}    & ICMPv6 Ping   \\
    \multicolumn{1}{|l|}{Port 7}      & \multicolumn{1}{l|}{ICMPv6 Ping} & \multicolumn{1}{l|}{ICMPv6 Ping}  & HTTP (80)     \\
    \multicolumn{1}{|l|}{HTTP (80)}   & \multicolumn{1}{c|}{-}           & \multicolumn{1}{l|}{HTTPS (443)}  & HTTPS (443)   \\
    \multicolumn{1}{|l|}{SSH (22)}    & \multicolumn{1}{c|}{-}           & \multicolumn{1}{l|}{Redis (6379)} & Port 65535    \\
    \multicolumn{1}{|l|}{HTTPS (443)} & \multicolumn{1}{c|}{-}           & \multicolumn{1}{l|}{6697 (IRC)}   & Telnet (23)   \\ \hline
    \end{tabular}
    }
    \caption{Temporal differences in protocols targeted by IPv6 scanning (as observed in previous background radiation/scanning studies)}
    \vspace{-1em}
    \label{tab:prevalence:temporal}
\end{table}

\section{IPv6 Scanner Address Discovery}\label{sec:discovery}

In this section, we focus on answering 
\emph{What address discovery strategies do IPv6 scanners use?} Put another
way, what types of observable activities monitored 
by IPv6 scanners to discover IPv6 scanning targets. 
%To answer this question, we 
%check for a causal relationship between service deployment 
%and IPv6 scanning activity. If such a relationship exists between them, it
%would imply that the scanner leverages information made available through
%the service as a means to discover a region for IPv6 scanning. We describe our
%methods  in
%\Cref{sec:discovery:methodology} and  our results in
%\Cref{sec:discovery:results}.

\subsection{Methodology}\label{sec:discovery:methodology}
\para{Overview.} We allocate and start each of our services
in a randomly selected group of four \subnet{64} subnets in a way that allows
us to identify and isolate any causal effects that the services may have on
IPv6 scanning (\Cref{sec:discovery:methodology:isolating}). Then, we use
a \emph{difference-in-differences} approach to test whether deploying
a particular service has a causal effect on IPv6 scanning activity
(\Cref{sec:discovery:methodology:causal}). Finally, we measure the duration of
any causal effects to uncover the typical lasting effect of deploying each of
the services used in our study (\Cref{sec:discovery:methodology:residual}).

\subsubsection{Isolating the effects of services}
\label{sec:discovery:methodology:isolating}
Our large network allocation allows us to run a controlled experiment in which
the `liveness' of specific \subnet{64} subnets of our network can be leaked (by
deploying services in them) for IPv6 scanners to find and the effects (\ie the
increase in scanning activity) of these leaks can be measured. %To
%improve our ability to correctly associate a measured effect (increase in
%scanning activity) with a specific leak (service), 
We construct our experiment
with the following key design decisions.

\para{Preventing mixing of effects.} Our experiment
    minimizes effects caused by two different
    types of services (\eg Tor and NTP pool) from co-occurring. Not taking
    steps to explicitly prevent this may lead to incorrectly associating 
    a causal relationship between a service and an observed effect.  
    Our strategy to achieve this is to ensure that, at any point in
    time, only one type of service is deployed in our entire \subnet{56}
    allocation. This reduces the probability of several simultaneous effects
    that are caused by different services from being mistakenly considered as
    a single effect caused by one service. 
    
\para{Erasing the impact of internal residual effects.} A residual effect in
    our setting is the increase in scanning activity that persists even after
    a service is discontinued. An internal residual effect is the effect that
    persists inside the \subnet{64} subnet that previously hosted a service. 
    For example, scanners that learned of the liveness of
    a \subnet{64} subnet because of a Tor relay operating in that subnet may
    continue scanning it even after the relay is turned off, causing a lasting
    effect within the \subnet{64} subnet. Not minimizing the impact of residual
    effects due to previously deployed services can harm the accuracy of 
    measured effect sizes for subsequently deployed services. 
    It is reasonable to expect that the residual effects of a service are
    likely to be stronger inside the \subnet{64} subnet that hosted it than
    elsewhere in our \subnet{56} allocation. Therefore, we ensure that no
    single \subnet{64} is used by multiple services for the entire duration of
    this study. This prevents internal residual effects from impacting
    subsequently measured effects.
    We have two sets of services that seem as an exception to this
    rule, but are not. First, the `OpenWPM' and `wget' crawlers are two
    techniques for the same service and so were deployed in the same treatment
    subnets; and what we refer to as the `NTP$_{pool}$' and `NTP$_{pool-20}$'
    NTP pool services are two different dates of the same NTP pool
    deployment --- one when the NTP pool servers were actually deployed and one
    when the deployed servers were marked as stable and available for public
    use. 
    
\para{Reducing the impact of external residual effects.} Residual effects from
    a prior service may also exist outside of its own \subnet{64} subnet. For
    example, a scanner that learned of the liveness of a \subnet{64} subnet
    because of a Tor relay operating in it may also scan neighboring
    \subnet{64} subnets for live hosts or services. Note that it is impossible
    to completely remove the impact of external residual effects --- after all,
    we cannot know exactly which other \subnet{64} subnets a scanner might
    target after learning of a service in one. Therefore, we can only make
    a best-effort attempt to reduce the impact of such residual effects on our
    subsequent measurements of effect sizes. We do this by
    creating an upper-bound for external residual effects which in turn makes our
    subsequent measurements of effect sizes conservative.
    We achieve this by running each of our services for an extended period of
    time (\emph{at least} two weeks). This allows any previously deployed
    services' external residual effects on the control region of our
    \subnet{56} to be reasonably integrated into the baseline (\ie mean number
    of scans in \subnet{64} control subnets) upon which future service
    deployments may be compared with --- thereby allowing for a conservative
    attempt at measuring future service effect sizes.

\para{Reducing the impact of random effects.} Since it is possible for scanners
    to chance upon one of our `live' \subnet{64} subnets, we reduce
    the impact of random effects that may be unrelated to our services. We
    do this by deploying a single service (\eg Tor relay) simultaneously on
    four randomly selected \subnet{64} subnets. Two of the selected \subnet{64}
    subnets are randomly chosen and have services deployed in the lower byte of
    their address space while the remaining two have service deployed at
    a random IP address contained in them. By measuring and
    reporting the average effect observed across all four of these subnets, we
    effectively reduce the impact of random effects.
\emph{These strategies improve: (1) our ability to
correctly associate increases in scanning activity with specific services and
(2) reduce the impact of latent confounding variables on the measured scanning
activities.}

\subsubsection{Identifying causal relationships and effect sizes}
\label{sec:discovery:methodology:causal}

Our efforts in \Cref{sec:discovery:methodology:isolating} allow us to
mitigate the impact of many latent confounding variables that may affect the
amount of scanning activity before and after a service is deployed, both
inside or outside a service's \subnet{64} subnets.
We use this controlled experimental setup to identify the
increase in scanning activity as a result of deploying a service. We measure
these effects using a difference-in-differences approach.

\para{Obtaining control and treatment groups.}
    We divide the 256 \subnet{64} subnets in our \subnet{56} address space
    into one \emph{control group} (\Cg) consisting of
    232 subnets in which no services were ever deployed and six
    \emph{treatment groups} (\Tg{crawl}, \Tg{dns}, \Tg{ntp-pool}, \Tg{ntp-public},
    \Tg{tor}, and \Tg{zone}) with each containing the four subnets allocated to the
    corresponding service. Recall, that no more than one of our treatment groups is
    active at any point in time and the intersection between any pair of treatment
    groups is null (\Cref{sec:discovery:methodology:isolating}).
%
%
%\begin{figure}[t]
%    \centering
%    \includegraphics[trim=125 0 95 120, clip, width=0.4\textwidth]{v6scan-diffs.pdf}
%    \caption{An illustration of the measured $\Delta_{s, t}^{\text{pre}}$ and
%    $\Delta_{s, t}^{\text{post}}$ for a service deployed at $t_s$ (as explained
%    in \Cref{sec:discovery:methodology:causal}). The corresponding measured
%    treatment effect size is \ddiff{s, t} = $\Delta_{s,t}^{post}
%    - \Delta_{s,t}^{pre}$.}
%    \label{fig:discovery:methodology:effectsize}
%\end{figure}

\para{Measuring narrow scanner effect sizes.}
    We define a narrow scanner as one which is focused on identifying active
    hosts within a specific \subnet{64} subnet. For such scanners, we expect to
    see an increase in scanning activity only within the treatment subnet and
    no effects in the control region. We measure their effect sizes using the
    techniques described below.
    Let $t_s$ denote the time at which the service associated with treatment
    group \Tg{s} is deployed and \meanscansperday{t_i, t_j}{$\mathcal{S}$}
    denote the mean number of scans observed per day between the days $t_i$
    and $t_j$ per \subnet{64} subnet belonging to group $\mathcal{S}$ (where
    $\mathcal{S}$ is the control group or one of our treatment groups). 
    Now, the mean per-day difference in scanning activity occurring in \Cg and
    \Tg{s} for the $t$ days prior to the start of a service $s$ (\ie the last $t$
    pre-treatment days) is: $\Delta_{s, t}^{pre} = (\text{\meanscansperday{t_s-t,t_s}{\Tg{s}}}
      - \text{\meanscansperday{t_s-t,t_s}{\Cg}})$.
    Similarly, the mean per-day difference in scanning activity occurring in \Cg
    and \Tg{s} for the $t$ days after the start of a service $s$ (\ie the first
    $t$ post-treatment days) is: $\Delta_{s, t}^{post} = (\text{\meanscansperday{t_s,t_s+t}{\Tg{s}}}
          - \text{\meanscansperday{t_s,t_s+t}{\Cg}})$
    Then, we measure the \emph{narrow scanner treatment effect size} over $t$
    days for the service $s$ as:
    $      \text{\ddiff{s, t}} = \Delta_{s, t}^{post} - \Delta_{s, t}^{pre} $.
    Described in words, \ddiff{s, t} is the difference of the differences
    between the mean per-day scanning rates between \Cg and \Tg{s} observed
    over $t$ pre-treatment and $t$ post-treatment days. An illustration of
    the measured treatment effect size is shown in the appendix
    (\Cref{fig:discovery:methodology:effectsize}). When \ddiff{s} is positive, it
    indicates increased scanning activity occurs inside the \subnet{64} subnets
    which contain the service $s$. When \ddiff{s} is negative, it indicates
    that scanner activity is focused on the region outside the \subnet{64}
    subnets --- \eg scanners may focus their attention on discovering other
    active subnets if they know a nearby subnet is live.

\para{Measuring broad scanner effect sizes.} We define a broad scanner as one
    which is focused on identifying active hosts in subnets that were not part
    of our treatment group. This is possible due to a different goal of
    identifying other live subnets rather than host IPs.
    Note that \ddiff{s} does not capture the increase in scanning activity
    caused by scanners which equally focus their attention on the control and
    treatment subnets. For example, if a scanner begins scanning our entire
    \subnet{56} subnet space in a uniform fashion after discovering one of our
    treatment services, $s$, \ddiff{s} will be measured as 0. Therefore, we use
    a simple difference method, applied over all our control subnets, to measure
    the effects of these scanners.
    Specifically, we measure the \emph{broad scanner treatment effect size}
    over $t$ days for the service $s$ as:
   % \begin{equation}
     $ \text{\dcontrol{s, t}} = \text{\meanscansperday{t_s,t_s+t}{\Cg}}
      - \text{\meanscansperday{t_s,t_s-t}{\Cg}}$.
    %\end{equation}
%
    Described in words, \dcontrol{s,t} is simply the difference in the average
    scanning activity observed before and after a service $s$ is deployed,
    computed over $t$ days, and over all control subnets in our
    \subnet{56} allocation.
    In the remainder of our reporting on the effect sizes \ddiff{s} and
    \dcontrol{s} in \Cref{sec:discovery:results}, we set $t$ as 14 days which is
    the duration of each service deployment.

\para{Verifying statistical significance of measured effect sizes.}
    To verify the statistical significance of the measured treatment effect
    sizes, we calculate bootstrap confidence intervals \cite{efron1987better,
    diciccio1996bootstrap} by resampling from the treatment and control subnets
    with replacement, using the percentile method to form a 95\% confidence
    interval with the 2.5\% and 97.5\% bootstrap percentiles on 10K samples. 
    We also calculate the $p$-values for a two-sided Welch's $t$-test
    \cite{welch1947generalization} to test for a null difference between (1)
    $\Delta_{s, t}^{pre}$ and $\Delta_{s, t}^{post}$ in the case of narrow
    scanners and (2) \meanscansperday{t_s,t_s+t}{\Cg} and
    \meanscansperday{t_s-t,t_s}{\Cg} in the case of broad scanners. 

\subsubsection{Measuring internal residual effects}
\label{sec:discovery:methodology:residual}
Since we do not reuse subnets in any of our treatment groups, we can
continue monitoring them for any residual effects that remain after the
service within it is no longer active. We measure these internal residual
effects by measuring $\Delta_{s}^{post}$ over a 14-day sliding window for
a period of 90 days after the service is halted. Lower values indicate a return
to normal scanning activity, and indicate that the \Tg{s}
subnets are received little to no additional scanning activity than the \Cg
subnets. This monitoring gives us an insight into scanner probing strategies
--- particularly, the freshness of their target lists.

% (1) fixing our $\Delta_{s}^{pre}$ to only reflect the difference in
% mean scans per day per subnet belonging to \Tg{s} and \Cg when considering the
% 14 days prior to service deployment --- \ie:
% \begin{equation}
%   \Delta_s^{pre}  = \text{\meanscansperday{t_s-14}{\Tg{s}}}
%   - \text{\meanscansperday{t_s-14}{\Cg}}
% \end{equation}
% and (2) calculating $\Delta_{s}^{post}$ over a sliding 14-day window
% starting at $t_{s'} = t_{s} + 14$, for a period of up to 90 days after service
% deployment ends --- \ie we compute:
% \begin{equation}
%   \Delta_{s, i}^{post} = \text{\meanscansperday{t_{s'} + i, 14 + i}{\Tg{s}}}
%   - \text{\meanscansperday{t_s + i, 14 + i}{\Cg}}.
% \end{equation}
% This allows us to compute the internal residual effect, $\Delta^{res}$ over
% time after the service is no longer active. Specifically, we compute:
% \begin{equation}
%   \Delta_{s, i}^{res} = \Delta_{s, i}^{post} - \Delta_s^{pre}
% \end{equation}
% We repeat our previous tests of statistical significance to test for the
% null difference between $\Delta_{s}^{pre}$ and each $\Delta_{s,i}^{post}$.
% We report the period of significant internal residual effect as the lowest $i$
% after which the differences are not statistically significant.
% 
\begin{table}[t]
  \centering
  \small
  \scalebox{\tabularscale} {
  \begin{tabular}{lcc|cc}%p{.9in}p{.9in}}
    \toprule
    \multirow{2}{*}{\bf Service}&\multicolumn{2}{c}{{\bf DNS logs}}&\multicolumn{2}{c}{{\bf PCAPs}}   \\
    
                    &  {\ddiff{s}}  & {\dcontrol{s}} & {\ddiff{s}} & {\dcontrol{s}}  \\
    \midrule
    wget            & {\bf 511.1}  &  {    -2.9}    & {0.0}       &  {0.0}            \\
    OpenWPM         & {\bf 564.0}  &  {    -0.1}    & {-0.4}      &  {0.4}            \\
    DNS probes      & {\bf 736.0}  &  {     1.9}    & {\bf 128.8} &  {\bf 265.1}      \\
    NTP$_{pool}$    & {\bf 348.3}  &  {\bf  6.0}    & {\bf 313.6} &  {\bf 734.4}      \\
    NTP$_{pool-20}$ & {\bf   6.5}  &  {\bf -9.7}    & {636.9}     &  {0.0}            \\
    NTP$_{public}$  & {\bf  72.2}  &  {     1.6}    & {\bf 116.3} &  {0.0}            \\
    Tor relay            & {\bf  87.1}  &  {    -0.4}    & {0.0}       &  {0.0}            \\
    DNS Zone            & {\bf   1.2}  &  {    -1.3}    & {-54.4}     &  {\bf 588.1}      \\
    \bottomrule
  \end{tabular}
  }
  \caption{Treatment effects observed in DNS logs and PCAPs.
  Services with a statistically significant \ddiff{s} or \dcontrol{s} ($p$
  < .05) are in {\bf bold}. \emph{Cf.} \Cref{sec:discovery:methodology:causal}
  for details on \ddiff{~}, \dcontrol{~}, and significance testing.}
  \label{tab:discovery:results}
\end{table}

\subsection{Results}\label{sec:discovery:results}

We first present the results of our causal analysis to identify treatment
effect sizes caused by IPv6 NXDOMAIN scanners
(\Cref{sec:discovery:results:dns-effects}) and all other IPv6 scanners
(\Cref{sec:discovery:results:ip-effects}). Next, we report the internal
residual effects caused by these scanners
(\Cref{sec:discovery:results:residue}).

\subsubsection{NXDOMAIN scanning treatment effects}
\label{sec:discovery:results:dns-effects}

The effects measured by the deployment of each of our services, on scanners
performing IPv6 NXDOMAIN scanning, are shown in the DNS logs columns of
\Cref{tab:discovery:results}. Interestingly, we found that all scanners
identified to be using the NXDOMAIN scanning approach may be categorized as
narrow scanners --- \ie they primarily focused their
scans inside our treatment subnets and had little to no scanning activity on
our control subnets (characterized by high \ddiff{s} and low
\dcontrol{s}). All services were found to have a statistically significant
\ddiff{~} effect, although varying widely in their magnitudes.

\para{Effects of direct contact with potential scanners.}
The largest \ddiff{~} effect sizes were observed by our DNS probes service
which sent DNS requests to a large number of IPv6 open resolvers. The effects
of this service ({\em cf.} \Cref{fig:discovery:results:dns-effects:dns} in
appendix) show an immediate rise in IPv6 NXDOMAIN scanning activity which
exclusively focusses on the subnets which sent the DNS requests
(\ie have low \dcontrol{~} values).
Our crawling services, wget and OpenWPM, have similarly large effects with
\ddiff{~} of 511 and 564 scans/subnet/day, respectively. 

\para{Effects of indirect contact with potential scanners.}
%We also illustrate the two services which cause our IP addresses to get
%added to archived public lists --- Tor relay and NTP public server in
%\Cref{fig:discovery:results:dns-effects:tor} and
%\Cref{fig:discovery:results:dns-effects:ntp-public}, respectively. 
We also find that the two services, Tor and NTP, which cause our IP addresses
to get added to archived public lists result in a delayed response from IPv6
scanners ({\em cf.} \Cref{fig:discovery:results:dns-effects:tor} and
\Cref{fig:discovery:results:dns-effects:ntp-public} in appendix). This
behavior is expected due to two reasons: First, both services
need participants to demonstrate a measure of reliability before they are
publicly listed in the consensus as usable for Tor and NTP users. Second,
unlike some of our other services (\eg the DNS probes and crawlers), we are not
directly interacting with potential scanners and instead are listing ourselves
at a place where scanners may be monitoring at a pre-determined frequency. 
Next, we see that the initial deployment of the NTP$_{pool}$ service results
in a significant increase in scanning activity with the \subnet{64} subnets,
but achieving enough reliability to be promoted into the public NTP pool only
results in a marginally increased scanning rate (an additional 6.5
scans/subnet/day greater than without the promotion). This is expected
since our servers' promotion within the NTP pool would be completely
inconsequential to scanners seeking to discover active IPv6 subnets.
Finally, we see that our entries in zone files had close to no impact on
the NXDOMAIN scanners with a \ddiff{~} of only 1.2.% (although statistically
%significant). 

% \begin{figure*}[t]
%   \centering
%   \begin{subfigure}[t]{.325\textwidth}
%     \includegraphics[width=\textwidth]{./plots/ate/ip-ddiff-ate-dns.pdf}
%     \caption{DNS probes}
%     \label{fig:discovery:results:ip-effects:dns}
%   \end{subfigure}
%   \begin{subfigure}[t]{.325\textwidth}
%     \includegraphics[width=\textwidth]{./plots/ate/ip-ddiff-ate-ntp-pool.pdf}
%     \caption{NTP pool server}
%     \label{fig:discovery:results:ip-effects:ntp-pool}
%   \end{subfigure}
%   \begin{subfigure}[t]{.325\textwidth}
%     \includegraphics[width=\textwidth]{./plots/ate/ip-ddiff-ate-zone.pdf}
%     \caption{Zone file listing}
%     \label{fig:discovery:results:ip-effects:zone}
%   \end{subfigure}

%   \caption{Treatment effect sizes from deployed services observed in PCAPs.
%   Only select services are illustrated (\emph{Cf.}
%   \Cref{sec:discovery:results:ip-effects} for complete results). Note that
%   linear interpolation has been applied to these visualizations for
%   smoothing.}
%   \label{fig:discovery:results:ip-effects}
% \end{figure*}

\subsubsection{IP scanning treatment effects}
\label{sec:discovery:results:ip-effects}

The effects of our service deployment on the number of scans observed directly
by our \subnet{56} allocation are shown in the PCAPs column of 
\Cref{tab:discovery:results}. Three of our deployed services (DNS,
NTP$_{pool}$, and NTP$_{public}$) have a large and statistically
significant \ddiff{~} effect while three have a large and
statistically significant \dcontrol{~} effect (DNS, NTP$_{pool}$, and Zone).
NXDOMAIN scanners focus not only on the treatment subnets, but also on the
subnets in the control region, suggesting an intention
to identify in-use subnets rather than host addresses.

\para{Effects of direct contact with potential scanners.} Unlike the NXDOMAIN
scanners, we find that our crawlers did not attract any attention from IP
scanners. Both, OpenWPM and wget had marginal \ddiff{~} and \dcontrol{~} values
of $\in [-0.4, 0]$. 
However, we find that our DNS probes to open resolvers generate increased
scanning activity ({\em cf.} \Cref{fig:discovery:results:ip-effects:dns} in
appendix). Interestingly, we note that the
increase in scanning activity is seen in both, the control and the treatment 
subnets (\ddiff is 128.8 and \dcontrol is 265.1). Further investigation shows
that scanners first identified and probed the treatment region for two days 
before returning with probes for addresses located in the lower bytes of all
our 256 \subnet{64} subnets ({\em cf.}
\Cref{fig:discovery:results:ip-effects:backformore-dns} in appendix). This
behavior is explained in detail in \Cref{sec:behavior}.

% \begin{figure*}[t]
%   \centering
%   \begin{subfigure}[t]{.325\textwidth}
%     \includegraphics[width=\textwidth]{./plots/ate/dns-comebackformore-heatmap.pdf}
%     \caption{DNS probes}
%     \label{fig:discovery:results:ip-effects:backformore-dns}
%   \end{subfigure}
%   \begin{subfigure}[t]{.325\textwidth}
%     \includegraphics[width=\textwidth]{./plots/ate/dns-residues.pdf}
%     \caption{NTP pool server}
%     \label{fig:discovery:results:residues:dns}
%   \end{subfigure}
%   \begin{subfigure}[t]{.325\textwidth}
%     \includegraphics[width=\textwidth]{./plots/ate/ip-residues.pdf}
%     \caption{Zone file listing}
%     \label{fig:discovery:results:residues:ip}
%   \end{subfigure}

%   \caption{IP probes sent to a \subnet{58} region of our allocation after
%   deployment of our DNS probe service (corresponding to the day 10-15 region of
%   \Cref{fig:discovery:results:ip-effects:dns}). White highlighted cells
%   indicate the subnets used for our DNS probe services (treatment group).}
%   \label{}
% \end{figure*}
% 
% 
% 
% \begin{figure}[t]
%   \centering
%   \includegraphics[width=.5\textwidth]{./plots/ate/dns-comebackformore-heatmap.pdf}
%   \caption{IP probes sent to a \subnet{58} region of our allocation after
%   deployment of our DNS probe service (corresponding to the day 10-15 region of
%   \Cref{fig:discovery:results:ip-effects:dns}). White highlighted cells
%   indicate the subnets used for our DNS probe services (treatment group).}
%   \label{fig:discovery:results:ip-effects:backformore-dns}
%   %\vspace{-.5em}
% \end{figure}

\para{Effects of indirect contact with potential scanners.}
%through participation in public network protocols and services.}
We find a significant uptick in IP scanning activity in the treatment subnets
after the deployment of NTP$_{pool}$ and NTP$_{public}$ experiments.
However, effect of the NTP$_{pool}$ treatment was not found to be
statistically significant since scanner attention was focused on
only one of our four treatment subnets. Our DNS, zone, and NTP$_{pool}$
experiments also generated significant effects in our control region with
\dcontrol{~} values of 265, 734, and 588, respectively. Same as before, we
found that scanners began their probes focused on the treatment subnets or
their neighbors before expanding to further away subnets.

\begin{figure}[t]
  \centering
  \begin{subfigure}[t]{.5\textwidth}
    \includegraphics[width=\textwidth]{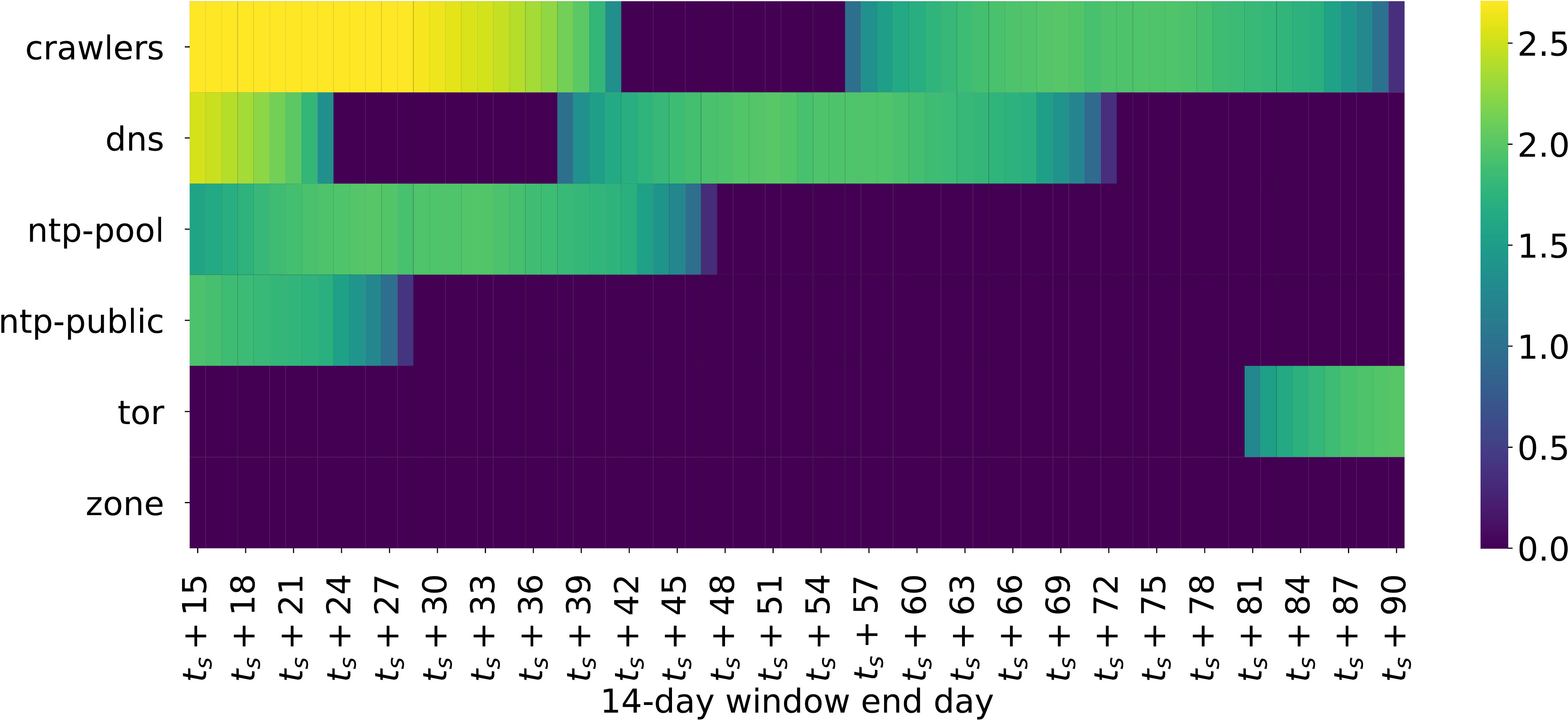}
    \caption{NXDOMAIN scanner residues ($\log_{10}$ scale)}
    \label{fig:discovery:results:residues:dns}
  \end{subfigure}

  \begin{subfigure}[t]{.5\textwidth}
    \includegraphics[width=\textwidth]{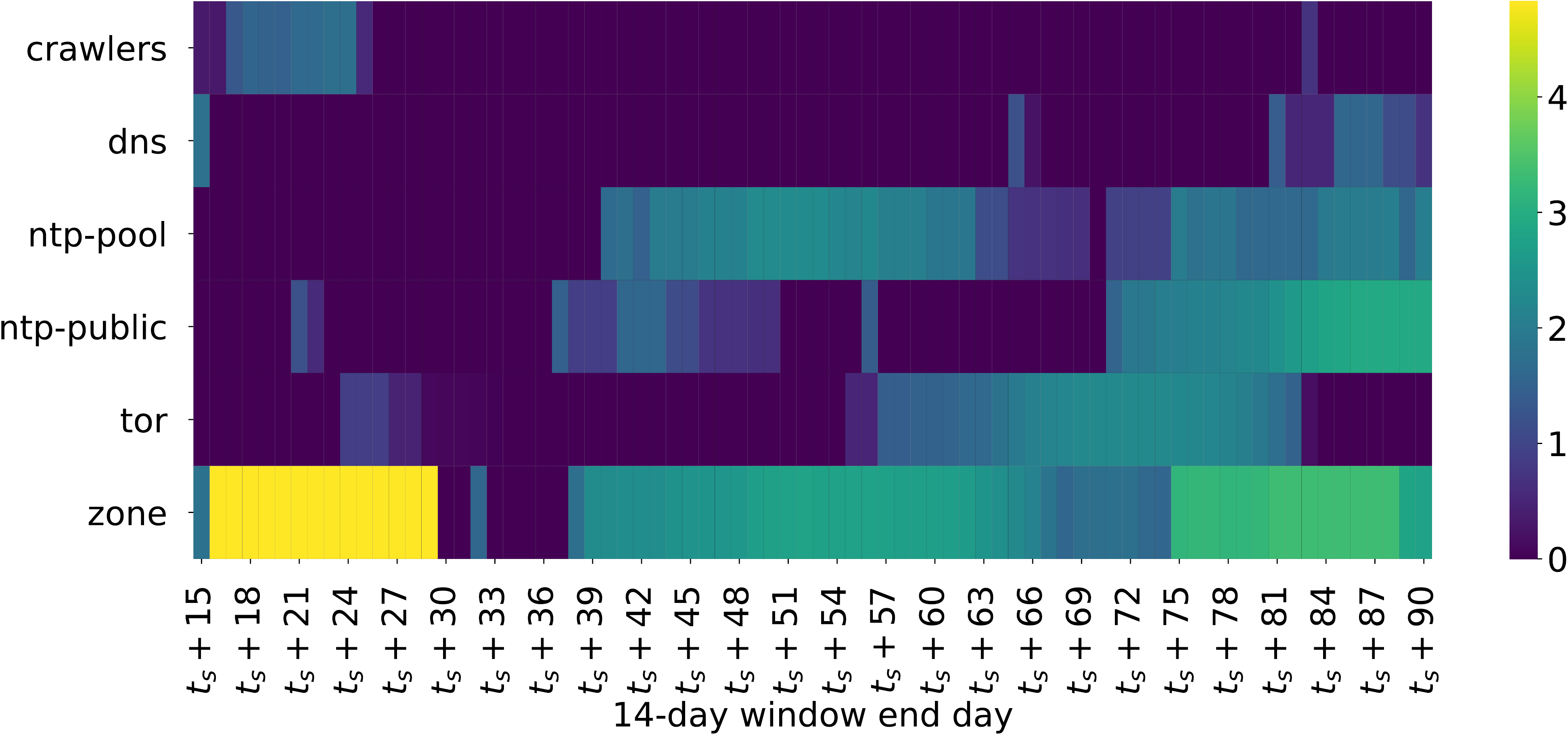}
    \caption{IP scanner residues ($\log_{10}$ scale)}
    \label{fig:discovery:results:residues:ip}
  \end{subfigure}
  \caption{Measured internal residue ($\Delta_{s}^{post}$) values over
  a sliding 14-day window starting from $t_s+15$ of each treatment until $t_s
  + 90$ days. 
  Higher values are indicative of focused scanning within the treatment subnet.
  Heatmap values are scaled by $\log_{10}$.}
  \label{fig:discovery:results:residues}
  %\vspace{-1em}
\end{figure}

\subsubsection{Internal residual effects}\label{sec:discovery:results:residue}
\para{Residual effects from NXDOMAIN scanners.}
\Cref{fig:discovery:results:residues:dns} shows the internal residual effects
within each of our treatment groups are high.
Specifically, we see that all subnets except for the ones used for Tor,
targeted by the NXDOMAIN scanners during their treatment period continue to be
scanned for a prolonged period of time --- in fact, our wget and OpenWPM
treatment subnets are targeted
by high daily volume of NXDOMAIN queries for two extended periods of
time separated only by a 15 day interval. This regularity and intensity is
interesting since these experiments aim to mimic a home network being
used for standard web browsing. We hypothesize that many scanners rely
on web traffic logs as data sources (as our crawler experiments might suggest)
to identify active hosts in such networks. 
We notice no NXDOMAIN scanning residue in the DNS, NTP$_{public}$, and
NTP$_{pool}$ treatment subnets 26-70 days after the start of
their treatments.

\para{Residual effects from IP scanners.} \Cref{fig:discovery:results:residues:ip}
shows the residual effects observed from IP scanners. Similar
to the NXDOMAIN scanners, we find prolonged IP scanning focused on our
treatment subnets for nearly all treatments with the exception of our direct
contact treatments --- the scans generated because of the crawlers and DNS
probes appear to end within 25 days and 1 day after our last use of their
treatment subnets, respectively. We find that our
listing in public sources of IPv6 addresses (NTP, Tor, and zone files) appears
to have a lasting residual effect on the treatment subnets. This is expected 
since their addresses are left to be discovered by scanners that eventually 
monitor archives of these lists.

\subsubsection{Takeaways} \label{sec:discovery:results:takeaways}

%We find that scanners leverage all of our tested methods to discover IPv6
%addresses. 
%
Our direct contact services (crawlers and DNS), which mimic user
activity in a home network, attract statistically significant
amounts of attention from NXDOMAIN scanners but not from IP probing scanners.
Services which contribute to public network protocols (NTP, Tor, and zone)
generate large amounts of scanning activity from one or both types of scanners. 
%
% Next, we stumble upon an unexpected finding:% that characterizes the use of each
%scanning approach. 
Unexpectedly, NXDOMAIN scanners never generate significant
amounts of traffic in the control subnets \emph{during the
treatment period}. This suggests that they are initially used to identify live
hosts within a \subnet{64} subnet rather than to identify other active
\subnet{64} subnets (low \dcontrol{s} values in
\Cref{tab:discovery:results}). In contrast, IP scanners are used
for both purposes (mixed \ddiff{s} and \dcontrol{s}
values in \Cref{tab:discovery:results}).
Finally, our study on residual scanning effects shows that the prior use of
a \subnet{64} subnet for specific purposes does impact the duration of residual
scanning activity. Notably, we find that our direct contact services,
which mimic home network activity, generate large and persistent volume of
NXDOMAIN scans but almost no IP scanning for a prolonged duration after the
subnet becomes inactive. On the other hand, hosting web services which result
in addition to the zone files generates consistent and significant volume of
IP scanning activity even 90 days after the service is
halted.

\section{IPv6 Scanner Behaviors}\label{sec:behavior}
% Our analysis in \Cref{sec:discovery} shows that, in general, scanner responses
% varied widely based on the type of scanning, how they
% discovered addresses, and whether they focused on the treatment or control
% regions of our address space. 
We now focus on the 
question: {\em Once scanners have discovered an active IPv6 subnet, what
behaviors do they exhibit?} 
% Specifically, we focus characterizations
% that let us understand how many scanners observed in the wild
% exhibit which types of behaviors (\eg lower-byte or random IID-focused scans, 
% narrow- or wide-area scans). We present our analysis of the behaviors
% observed in traditional IP scans (\Cref{sec:behavior:ip}) and then focus on
% NXDOMAIN scans (\Cref{sec:behavior:nxdomain}).

\subsection{IP scanning} \label{sec:behavior:ip}
\para{Overview.} We seek to 
answer three questions: (1) what is the dominant scanning strategy
for target address generation (\Cref{sec:behavior:ip:address-gen}) --- do
scanners target lower-byte or random addresses? (2) does the discovery of
a treatment subnet with an active end-host having a specific type of address
(lower-byte or random IID) impact the amount of scanning activity the subnet
receives? (\Cref{sec:behavior:ip:treatment-impact}) and (3) how does scanner
behavior change in scope over time --- \eg do they start as narrow-scanners and
expand to broad-area scanning? (\Cref{sec:behavior:ip:time})

\subsubsection{Target address generation strategies}
\label{sec:behavior:ip:address-gen}

\para{IP probe categorization.} 
To understand scanner address generation strategies, we grouped
each probe into two categories; probes targeted at lower-byte IID addresses
(\plb) and  probes targeted at random IID addresses (\prb). We placed a probe
in \plb if the IID section of the destination address (bits 64-128 in our case)
have \emph{at least} 40 leading zeros. 
If the probe did not meet this condition, we automatically assign it
to \prb. Therefore, all probes belong in either \plb or \prb. Although this
categorization appears simplistic, we settled on it only after noticing that it
was indeed the case that all probes from a scanner that were placed in \prb
were pseudo-random at the nybble-level (\ie had maximum entropy for the nybbles
they varied indicating that if a scanner altered higher nybbles, they tried all
values for that nybble with nearly equal probability). 
% This is illustrated in
% \Cref{fig:behavior:ip:heatmap-random} which shows the nybble frequency for a
% typical random IID scanner. 

% \begin{figure}[t]
%     \centering
%     \includegraphics[width=0.5\textwidth]{plots/char_plots/hm_rand.png}
%     \caption{Nybble frequencies for a typical random IID scanner.
%     \rnnote{label the axes, make text larger, and make PDF}}
%     \label{fig:behavior:ip:heatmap-random}
% \end{figure}

\para{Which addresses are the most common scanning targets?} 
In all, we observed 12.2M (84\%) probes from {696 (65.1\%)} scanners
addressed to {4,359} unique lower-byte addresses. In
comparison, we observed only 2.3M (16\%) probes from {747 (70\%)} scanners
addressed to {2,291} unique random IID addresses. {\em The average
lower-byte IID address was the target of 347,625$\times$ more scans than the
average random IID address.}

\para{Which scanning strategy do scanners most frequently use?} 
693 (65\%) of all observed scanners used an exclusive probing strategy --- \ie
either only probing lower-byte IIDs or only probing random IIDs. Of these, 321
were exclusively lower-byte scanners and the remaining 372 were exclusively
random IID scanners. The lower-byte only scanners generated only 4\% of all
probes, the random IID scanners generated only 5\% of all probes, and the
remaining 91\% of probes were generated by scanners using a mix of both. 
%
% \emph{Taken all together, we see that scanners generally
% focus their attention on lower-byte addresses. However, a majority of those
% observed (and in fact the largest scanners) do not exclusively probe only
% lower-byte addresses.}

\subsubsection{Impact of addresses of known active end-hosts}
\label{sec:behavior:ip:treatment-impact}

\begin{figure}[t]
    \centering
    \includegraphics[width=0.5\textwidth]{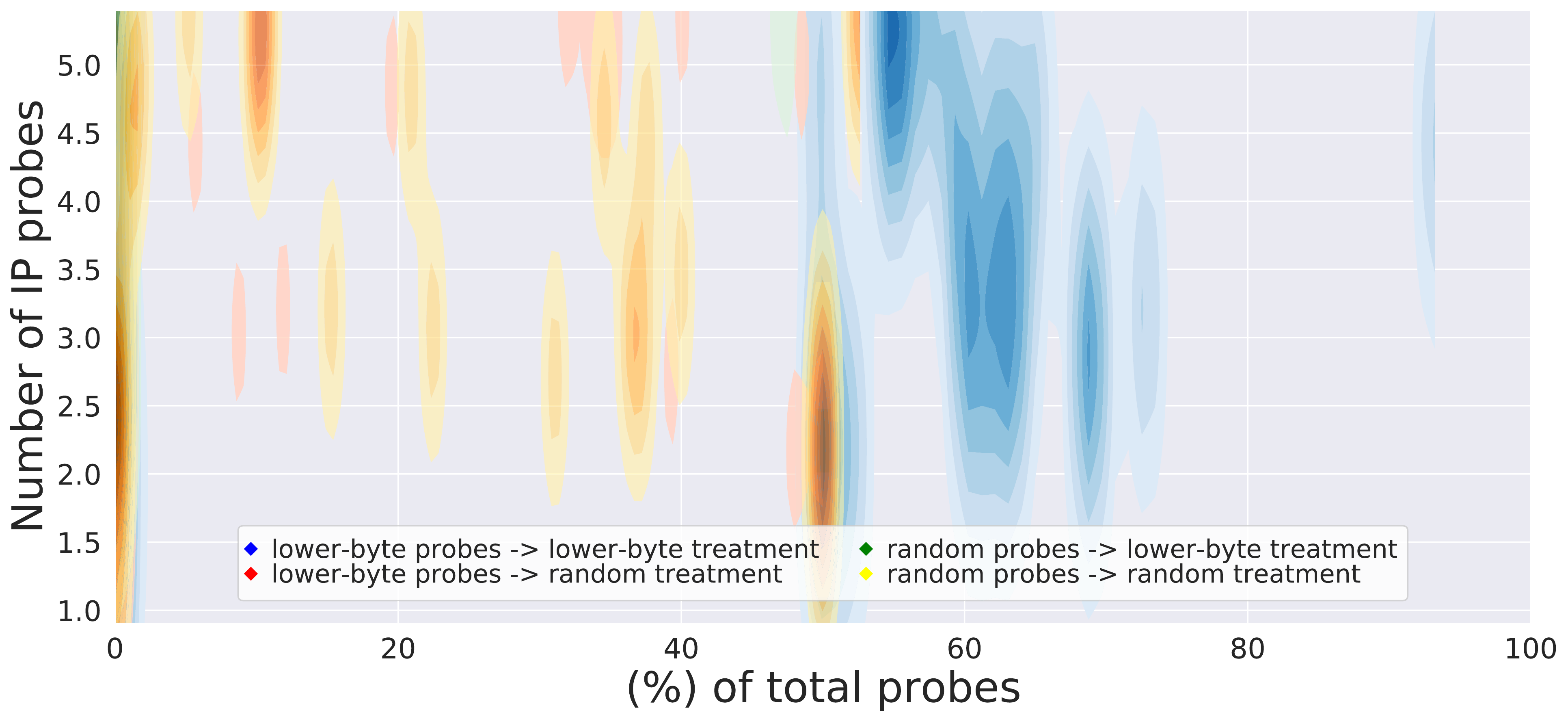}
    \caption{Types of probes sent to each type of treatment subnet by scanner
    sizes. ({\em Cf.} \Cref{sec:behavior:ip:treatment-impact}). Values are
    $\log_{10}$ scaled.}
    \label{fig:behavior:ip:treatment-impact}
\end{figure}

\para{Probe and destination subnet categorization.} 
To understand if scanner behavior depends on the locations of the known
end-hosts, we categorize each of our 24 treatment subnets into
\emph{lower-byte} (\slb) or \emph{random} (\srb) based on whether the services
deployed in them were on a lower-byte or random IID. We then categorized
each probe sent to a treatment subnet into four
buckets: \plb $\rightarrow$ \slb, \plb $\rightarrow$ \srb, \prb $\rightarrow$
\slb, and \prb $\rightarrow$ \srb.
\Cref{fig:behavior:ip:treatment-impact} shows how
scanners of different sizes probe these subnets. Here, a point 
indicates that scanners that sent a total number of probes corresponding to the
$y$ axis sent $x\%$ of their probes to a treatment subnet of the type indicated
by the color. The density shows the number of scanners with similar $x$, $y$.

% 256 \subnet{64} subnets as either: a {\em
% control (other)}, {\em neighbors}, {\em lower-byte treatment}, or {\em
% random treatment}. Assignments into the lower-byte and random treatment groups
% were made based on whether the service within the subnet was assigned
% a lower-byte IID or a random IID. There were 12 subnets in each of these
% treatment groups. Subnets that were adjacent to either of the treatment groups
% but not contained in either were placed in  the control neighbors group and all
% other subnets were placed in the control (other) group.
% 
\para{Do scanners change their scanning behavior based on known host
addresses?} {\em In short, yes.} We focus on the most dense
region of \Cref{fig:behavior:ip:treatment-impact} --- \ie the region with $x
\in [35, 70]$. We make two observations that lead us to the above
conclusion. First, scanners of all sizes send lower-byte
probes to lower-byte treatment subnets (indicated by the blue density region)
about 60-80\% of the time. Second, when presented with a subnet in which they
are aware of a random IID host, they send random probes (indicated by the
yellow density region) nearly 50\% of the time --- about 3$\times$ more than
the base rate. Further, no scanner ever sends random probes to
a subnet in which the presence of a lower-byte host is known (the only green
density region is at $x = 0$). For comparison, the base rate of random IID
probes is 16\%.
\begin{figure}[t]
    \centering
    \includegraphics[width=0.4\textwidth]{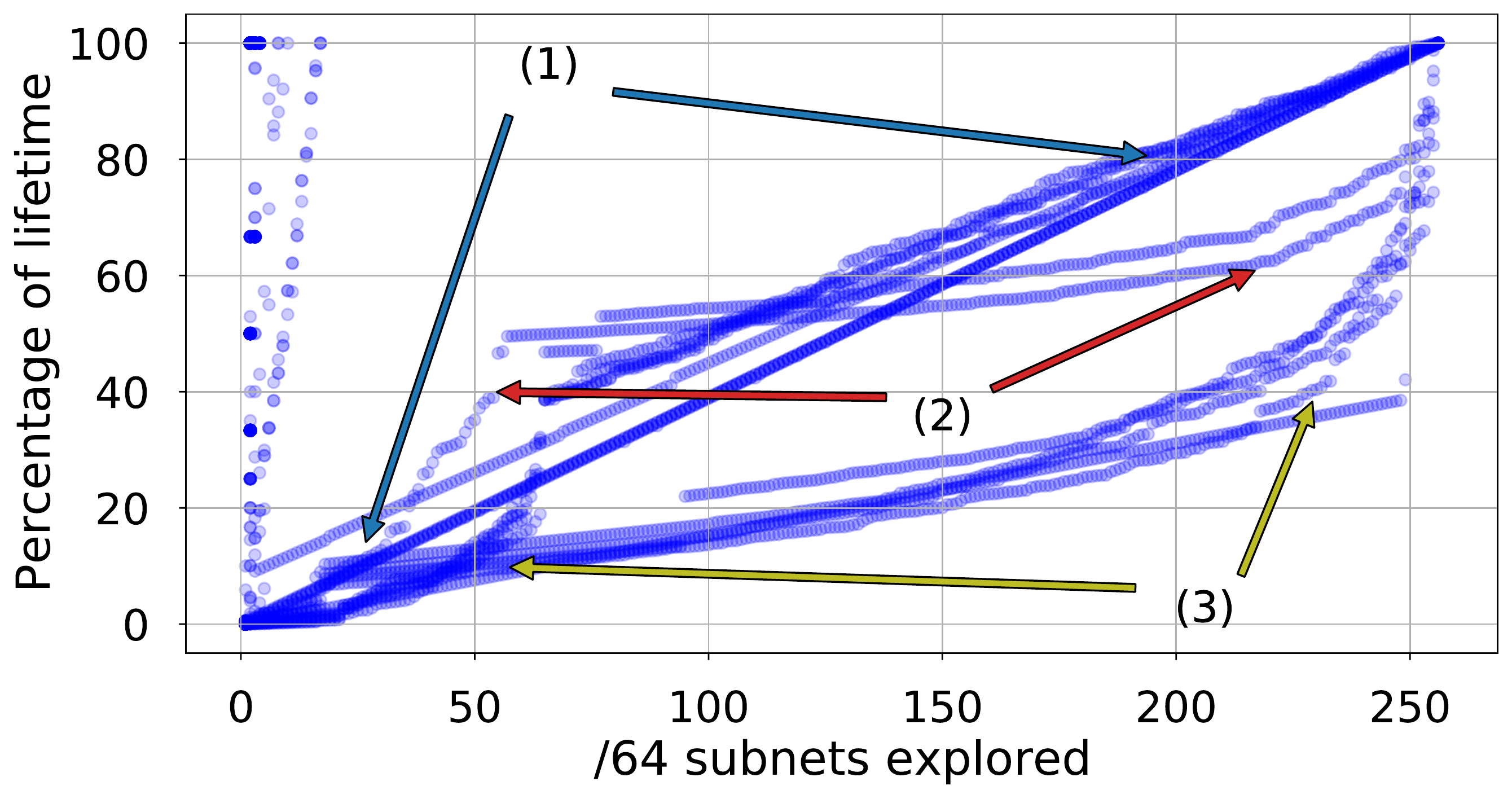}
    \caption{Subnets scanned over time by each IP scanner (\Cref{sec:behavior:ip:time}).}
    \label{fig:behavior:ip:time}
\end{figure}

\subsubsection{Change in scanner scope over time}
\label{sec:behavior:ip:time}

\para{Characterizing behavior by time.} To understand how scanners
change their scope over time, we analyze the number of \subnet{64} subnets
scanned by each scanner in their scanning lifespan (measured as
the fraction of their total probes sent until that point in time).
\Cref{fig:behavior:ip:time} illustrates the breadth of scanning behavior over
the lifetime of each scanner in our dataset. Each `dot' in an ($x$, $y$)
coordinate in the plot represents one point in the lifespan of one of the
scanners and indicates that there was a scanner that,
after having sent $y\%$ of their total probes ever sent, had probed $x$ unique
subnets. 

\para{Do scanners change their scope over time?}
\emph{In short, sometimes.} From \Cref{fig:behavior:ip:time}, we 
observe four trends. First, on the top left we find that the
narrow IP scanners do not change their
behavior and focus on a handful of subnets over their entire lifetime. Then, we
have one of three distinct behaviors that are observed in each of the broad IP
scanners. These are annotated with (1), (2), and (3). The scanners exhibiting
behavior (1) focus their scans equally on all our 256 subnets ---
sending the same number of probes in one before moving on to the next. 
These probes are mainly focused on the lower-bytes of each
subnet. Scanners that form points highlighted by (2) initially focus their
attention on a small number of subnets and then spread their probes broadly
amongst the remaining. Nearly 40\% of all their initial probes
appear in just 50-60 unique subnets (which make up less than 25\% of our
address space) and then the remaining 60\% of their probes sent later in time
are distributed across the remaining ~75\% of subnets. Scanners making up the
points marked by (3) show the opposite behavior in a more extreme form. 
They perform broad scans during the first part of their lifespan as they
probe more than 200 unique subnets with just the first 40\% of their probes
before they focus their attention on the subnets that remain.

\subsubsection{Takeaways}

We see diverse scanning behavior in some dimensions and
surprisingly uniform behavior in others. Our probe characterization
shows that probes are most likely to focus their attention on lower-byte
addresses in a subnet, but do not rule out probing random addresses.
In fact, we see that probing behavior changes to incorporate more random scans
in subnets where the presence of random addressing is already known to
a scanner, while this never occurs when the presence of a lower-byte host is
known. Finally, we see that there are just three distinct patterns
observable in IP scanners that perform broad scanning --- either uniformly
broad scanning, or deep scanning for the first 40\% of their probes followed by
broad scanning with the remaining 60\%, or vice-versa. We find that
all scanners either focus on a very small number of subnets or perform scanning
over all 256 subnets --- never in the middle.

\subsection{NXDOMAIN scanning}
\label{sec:behavior:nxdomain}

As discussed in \Cref{sec:prevalence:methodology}, \nx scanners can reduce the
search space of finding active hosts significantly. This is achieved through
scanning the \textit{ip6.arpa} reverse DNS tree at different subnet length
levels and discarding all subnet length trees under the current node if an
{\tt\nx} response is received --- after all, it is known that no domains exist
within that entire subnet. This is the reason why our results in
\Cref{sec:discovery} shows that there is little to no \nx scanning activity
outside our treatment subnets. This property makes it necessary for us to use
different metrics to characterize \nx scanning. Therefore, we focus our
analysis on understanding: when \nx scanners know of the presence of a host
with a particular address type (lower-byte or random) how do they scan the
subnet to identify other hosts --- do they assume lower byte nybbles at each
level or do they branch diversely and does this change depending on the subnet
level indicated in their probe?

\para{Characterizing probe types in NXDOMAIN scanning.}
Probes generated by \nx scanners can be of different lengths compared to
probes generated by IP scanners which are always full 128-bit addresses. To
understand probe generation techniques by \nx scanners, we divide our probes
into two categories; \lb and \rnd.  \lb probe of length $n$ is 
a reverse PTR query for a subnet of \subnet{n} that ends in 0. All non-zero
queries are \rnd probes.

\begin{figure}[t]
    \centering
    \includegraphics[width=0.5\textwidth]{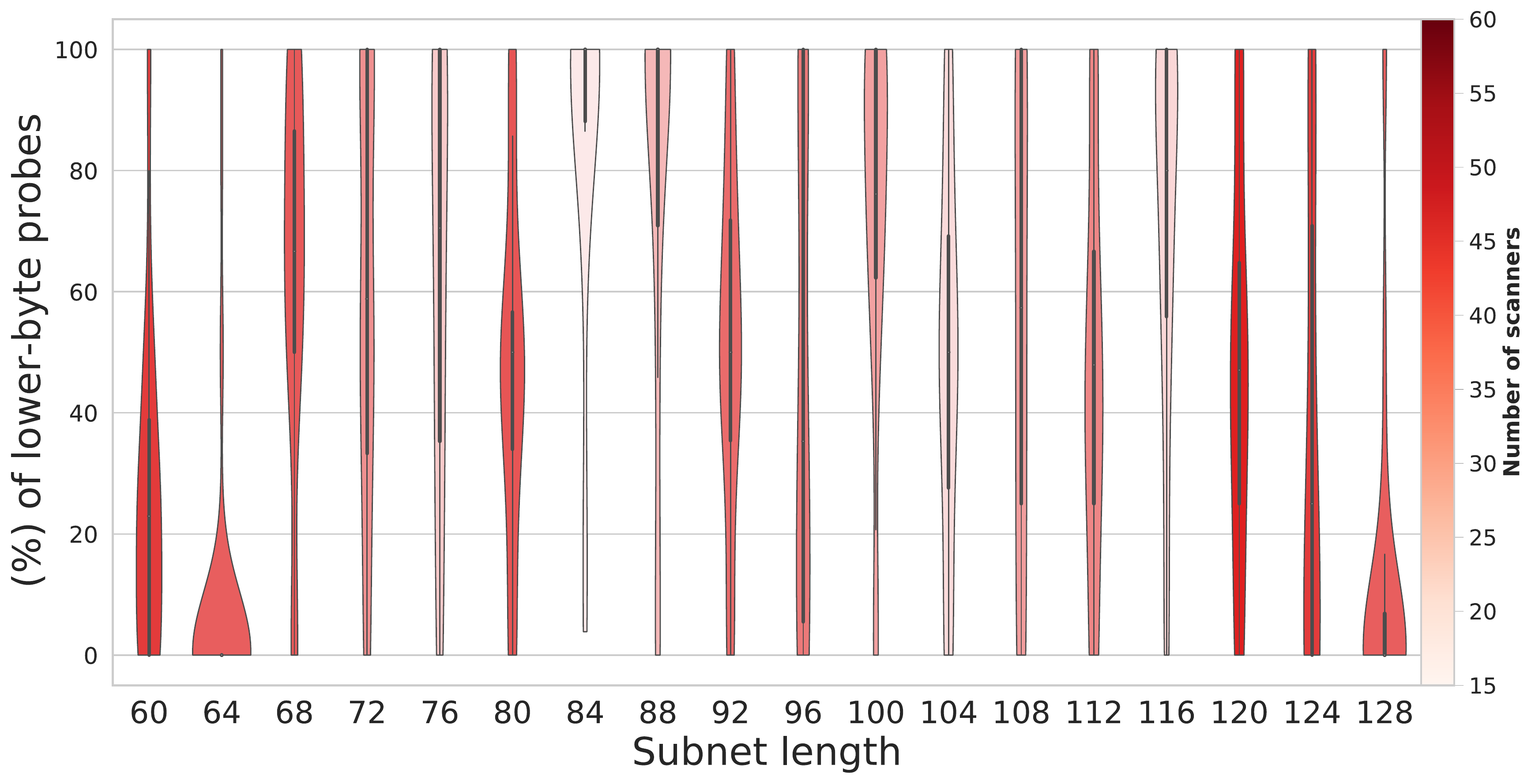}
    \caption{\% of \lb NXDOMAIN scans at each subnet length.}
    \label{fig:behavior:nxdomain:violin}
\end{figure}

\para{NXDOMAIN scanning by probe type and subnet length.}
\Cref{fig:behavior:nxdomain:violin} shows: (1) the number of scanners that sent
probes for a subnet of the specified length and (2) whether the probes at this
level were for the `0' subtree (\ie a lower-byte probe) or `1-f' subtree (\ie
a random probe). Darker colors associated with the
violin of a subnet length indicate more scanners sent probes at that subnet
level. We find that most scanners focus their attention on \subnet{60},
\subnet{64}, \subnet{80}, \subnet{120}, and \subnet{124}. This suggests that
larger number of subnets expect hosts to be allocated within these subnet
ranges. 
% Since a \subnet{64} IID allocation is only recommended and not
% expected, one might also consider these various levels of subnets a measure of
% the  scanners' expectation of the possible prefix lengths. 
We now analyze the distributions (depicted by the violins) of lower-byte and
random probes at each subnet level. The width of each violin is indicative of
the number of scanners and that have  $y\%$ of lower-byte probes (as a fraction
of all their probes at that subnet level). Violins with wide areas at the
bottom (\eg \subnet{64} and \subnet{128}) suggest that scanners conduct mostly
random scanning when sending queries at that level. On the other hand, smaller
lower areas indicate more propensity for lower-byte scanning. We see that with
the exception of the \subnet{64} and \subnet{128}, even NXDOMAIN scanners
mostly rely on lower-byte probes (or have an expectation of host deployments at
lower-byte addresses).

\section{Security implications for network operators}\label{sec:security}

In this section we discuss the implications of our findings 
for operators of IPv6 networks on the Internet.

\para{Implications for address assignment in networks.}
We show that a lower-byte IID IPv6 address receives 347,625$\times$ the scans
received by a random IID address (\Cref{sec:behavior:ip:address-gen}). 
Moreover, ~99\% of all IPv6 probes in our measurements
targeted a lower-byte address. Therefore, we recommend
network operators to use semantically opaque interface
identifiers \cite{RFC7217} for assigning IPv6 addresses to hosts
in their networks. Such an address assignment will increase the
scanning overhead by forcing scanners to probe $2^{64}$ addresses 
in the worst case and reduce the volume of probes received by an 
individual host.

\para{Implications for allocating scanning defense budgets.}
We show that the volume of IP probe-based scanning activity, 
which is threatening due to direct interaction between 
the payloads and potentially vulnerable hosts, attracted by a 
subnet is highly dependent on the type of host activity 
in the network. This allows operators to consider allocating 
a security budget for their network that uses the expected 
host activities and addressing schemes in use to estimate 
scanning activity.

\para{Implications for IPv6 subnet reuse.}
Our measurements show that operators can safely reuse subnets 
for different hosts in special cases without the risk of 
attracting scanners that targeted the previous host of the subnet (\Cref{sec:discovery:results:takeaways}).

\para{Implications on the utility of threat exchanges.}
Security practitioners have believed that increasing adoption of IPv6
on the Internet would reduce the the utility of Internet threat exchanges
due to the infeasibility of blocklisting in the vast IPv6 address space~\cite{Spamhaus98:online, WillIPv694:online}.
Blocking at the granularity of \subnet{64} subnets increases the 
feasibility of blocklisting-based approaches with the potential of
collateral damage due to hosts sharing fate on the same \subnet{64}
subnet~\cite{li2020towards}. We show that there is value 
in using blocklists for IPv6, at least in the current IPv6 
landscape (\Cref{sec:prevalence:characteristics}). Majority
of the scanning traffic observed by our subnet originated from IPv6
addresses already listed on popular threat exchanges like AbuseIPDB.
Most of the scanners not present on the blocklist belonged to a 
Research/Education AS. Other scanners that sent significant 
scanning traffic while not being reported on the IP blocklist
belonged to either a \subnet{64}, \subnet{48} or a \subnet{32} 
shared by a scanner that was previously reported on the blocklist. 

\para{Implications for generating DNS PTR Records.}
The NXDOMAIN scanning technique (\Cref{sec:background}) can significantly reduce the number
of probes required to find an active host in an IPv6 subnet.
Thus, NXDOMAIN scanning can potentially bypass the security that
semantically opaque identifiers would provide from active scanning. 
Dynamically generating PTR records when queried ("On the fly")
as described in RFC 8501 \cite{RFC8501} can render this scanning technique
ineffective. We recommend network operators to use this technique since it enables
generation of a valid PTR record for every address that is queried, 
preventing the scanner from exploiting the NXDOMAIN semantic to discard entire 
subtrees of addresses.

% \textbf{Generating PTR records \textit{"On the Fly"}.}: The NXDOMAIN
% scanning technique, as described in \Cref{sec:background} can reduce the number
% of probes required to find an active host significantly. In the worst case, the
% number of probes can be reduced from from $2^{64}$ to $16 * 64$ (to find a host
% within an interface identifier of 64 bits). This technique is also agnostic of
% the type of address and works equally well for lower-byte and random IID
% addresses. Therefore, NXDOMAIN scanning can potentially bypass the security that
% semantically opaque identifiers would provide from active scanning.  \\\para{\textit{Security
% consideration:}} Dynamically generating PTR records when queried ("On the fly")
% as described in RFC 8501 \cite{RFC8501} can render this scanning technique
% useless. If this technique is used, a valid PTR record would be returned for
% every address that is queried, thereby not allowing the scanner to exploit the
% NXDOMAIN semantic to discard whole subtrees.

\section{Related Work}\label{sec:related}
% Although there are many influences on our work, this research largely benefited
% from prior art that can be broken down in three dimensions: (1) address discovery
% research which aims to identify sources of IPv6 addresses for scanners and (2)
% address assignment measurement which aims to identify practices being used for
% address allocation in the wild, and (3) IPv6 scanning target generation
% research which seeks to develop techniques to generate candidates for IPv6
% probing based on seed live addresses. 
In this section, we highlight the influence of previous research efforts
on our study design and put our work in context.
%
% Next, we put our research in the context of other efforts seeking to shed light
% on scanner behavior.

\para{IPv6 address discovery.} Our decision to leak addresses through
direct contact and indirect contact methods was informed
by previous work by Gasser \etal \cite{gasser2016scanning}. Their work
uncovers sections of the live IPv6 address space leveraged a variety of address
sources.
%passive tap
%installed on an IXP, measurements of IPv6-capable web services, reverse-lookups
%of the entire IPv4 address space, and traceroutes to already known addresses.
%In total, they identified 150 million `live' IPv6 addresses from 82\% of all
%IPv6 Autonomous Systems. 
In their follow-up work \cite{gasser2018clusters,
IPv6Hitl95:online}, they emphasized the importance of focusing address
gathering on a wide range of sources and maximizing address stability
rather than raw count when measuring efficacy of address gathering approaches. 
Our decision to include DNS logs in our data logging framework was
informed by Fiebig \etal \cite{fiebig2017something, IPv6Farm87:online,
fiebig2018rdns} who were the first to leverage the NXDOMAIN scanning approach
described in \Cref{sec:background} to identify unknown active IPv6 addresses.
Their research has been incorporated
into off-the-shelf IPv6 scanners. Expanding on the work of Fiebig \etal,
Borgolte \etal \cite{borgolte2018enumerating} use DNSSEC-signed reverse-zones
to collect active IPv6 addresses and find it to work better than the NXDOMAIN
approach for scanning. 

\para{Measurement of address allocation practices and IPv6 target generation.}
In our study, we deployed services with
lower-byte and pseudo-random IIDs. This decision was inspired by
the work of Gont \etal \cite{RFC7707} who discussed patterns
observed in IPv6 addresses collected from web servers, mail servers, clients
and IPv6 routers observed in the wild. Their work showed that for each of these
different categories, a majority the of IPv6 addresses follow a pre-defined
address pattern --- specifically lower-byte IIDs and pseudo-random IIDs. 
% Some of these patterns include \textit{low-byte addresses}
% which was found in 92\% of mail server addresses and 70\% of router addresses.
% They also find that \~70\% of client addresses have a pseudo-random IID. 
% They
% argue that such common allocation patterns effectively reduce the IPv6 address
% space to a size that is feasible for an exhaustive scan.
%
Focusing on measuring the stability of assigned addresses, Plonka \etal analyze
active client addresses logged by a CDN \cite{plonka2015temporal}. They report
that the IIDs of discovered addresses are unstable, but the \subnet{64} subnets
associated with them are stable. 
% Additionally, they also discover the
% prominence of lower-byte IID addressing.
% % 
% although < 4\% of IPv6 addresses remain stable over a period of 14 days
% and 80\% of \subnet{64} subnets remain stable during the same time frame.
% Similar to Gont \etal, they find many IPv6 subnets using sequential lower-byte
% addressing. 
%
Along similar lines, Padmanabhan \etal conduct a temporal and spatial
comparison of IPv4 and IPv6 addresses using 3000 RIPE Atlas probes
\cite{padmanabhan2020dynamips} and find similar results regarding subnet
stability. 
% They find that IPv6 network subnets can remain
% stable for months for residential subscribers. Furthermore, they find that IPv6
% addresses remain stable for longer when compared to IPv4 addresses and that
% address reassignments in IPv6 occur inside the same BGP prefix.

\para{IPv6 target generation.}
The findings from the above research on allocation practices have a direct
impact on target generation algorithms used with IPv6 scanners. Foremski
\etal leverage information-theoretic entropy at the nybble level to
\textit{mine segments} of the IPv6 address with similar entropy
\cite{foremski2016entropy}. They apply this technique on addresses collected 
from a range of public data sources and find that client addresses have the
most entropy in the last 64 bits whereas servers tend to use lower-bit
addresses. 
% These techniques were then applied on a training set of 1000 IPv6
% addresses to generate 1 million candidate target addresses which resulted in
% the discovery of 770k new addresses and 46k prefixes.
%
Murdock \etal proposed \textit{6Gen} \cite{murdock2017target} --- an IPv6
target generation algorithm which uses a seed of input addresses to identify
dense address space regions and determine patterns present in the seed
addresses based on a {probe budget}. They find 56.7 million addresses in
aliased regions and 1 million addresses in non-aliased
regions. 

\para{Understanding scanner behavior.} Ours is not the
first work to shed light on the behavior of IPv6 scanners. In fact, these
efforts date back to 2006 when Ford \etal \cite{ford2006initial} conducted the
first measurement of IPv6 background radiation inside a darknet and found no
evidence of scanner-related traffic. 
In 2010, Huston and Kuhne \cite{huston2010background} repeated a similar
experiment and discovered only a few
thousand packets that were attributable to scanners (such as TCP SYN packets)
and generally observed traffic from common IPv6 misconfiguration. 
A  similar study was conducted by Czyz \etal \cite{czyz2013understanding} in
2013. They conducted a massive-scale measurement of IPv6 scanning by gathering
and analyzing unclaimed traffic received by five \subnet{12} IPv6 subnets.
Despite their large vantage point into IPv6 scanning behavior, their study also
only identified trace amounts of IPv6 scanning.
% yielded similar results as Huston and Kuhne --- \ie they discovered only small
% traces large-scale malicious scanning and instead found that observed traffic
% was mostly due to bad server configuration and routing instability. 
%
More recently, in 2018, Fukuda and Heidemann \cite{fukuda2018knocks} proposed
using DNS backscatter to identify IPv6 scanning activity. Of note is the fact
that this was the first study to find evidence of widespread IPv6 scanning
activity. By analyzing the logs of an authoritative DNS server for reverse-DNS
lookups, they identified (on average) 16 active IPv6 scanners per week over
a six month period. 
%
% In the IPv4 realm, Durumeric \etal \cite{durumeric2014internet} leveraged dark
% traffic from Merit network to study the infrastructure being leveraged by IPv4
% scanners. 
% \parait{IPv4 scanning behavior.}
% Although focused on uncovering the behavior of IPv4 scanners, Durumeric \etal
% \cite{durumeric2014internet} also leveraged dark traffic to study the
% infrastructure used by IPv4 scanners. 

The work of Richter and Berger
\cite{richter2019scanning}, while focused on IPv4 scanning behavior, informed
our decision to conduct a controlled experiment with specific address leaks.
Using a CDN as a vantage point, they find that different services
attract scanners with different scanning strategies. 
Durumeric \etal \cite{durumeric2014internet} also observe
similar patterns. They find that the services that scanners target change over 
time in IPv4; something we do not observe in IPv6. Like us,
they also find that cloud service providers are the source of most of the scanning
traffic.

% \parait{Our work in context.}
While similar in goal, our study differs significantly from the highlighted
IPv6 studies in two aspects: (1) IPv6 is much more ubiquitous today with nearly
35\% global adoption than the time at which these studies were conducted
(between 0.1\% to 3\%) \cite{GoogleIPv6} and (2) we take a unique approach by
conducting a controlled experiment to specifically identify the behavior of
scanners in response to specific types of services deployed in subnets
(\Cref{sec:discovery}) and types of addresses allocated (\Cref{sec:behavior}).

\section{Conclusions} \label{sec:discussion}

We study IPv6 scanner address discovery strategies and 
the behavior of scanners when information about host deployment 
is known to them. Our controlled experiments reveal several key
findings. First, we see different levels of focus and magnitude in the scanning
that occurs in response  to different types of address leakage --- some
resulting in the leaked subnets receiving, on average, 700 more
probes/day/subnet
while others result in all nearby subnets receiving large amounts in scanning.
Next, we see that many scanners change their scanning behavior when
a host in the subnet is already known to have a specific type of address.
However, lower-byte scanning still remains the dominant scanning strategy even
in subnets known to allocate random-IIDs to hosts. This results in lower-byte
addresses receiving over 300K times more probes than random addresses, on
average. Besides the key takeaway that {\em allocating lower-byte IIDs to
end-hosts results in significantly increased risks because of scanning}, this
study provides network operators with actionable insights regarding address
allocation strategies to mitigate the impact of scanning and what to expect from 
scanners when a subnet is expected to run specific types of services.
% The primary limitation of our work,
% is that we can only mitigate the effects of latent confounders on 
% our causal inference. However, such challenges are common in causal 
% inference studies conducted in natural
% scenarios such as ours and our efforts to mitigate them were carefully
% constructed prior to experimentation. 

% \section*{}
\textbf{Acknowledgments.} We would like to thank the the ITS and CLAS Linux groups at the University of Iowa (JJ Urich, Jay Ford, Scott Allendorf, Matthew Brockman, Hugh Brown, Bradley Carson, Aran Cox, and Dan Holstad) for their help in setting up the experimentation testbed for this study. We would also like to thank our reviewers for their valuable feedback and suggestions. This research was funded by the National Science Foundation under award \#1953983. The opinions, interpretations, conclusions, and recommendations pare those of the authors and are not endorsed by the funding bodies.

%\balance

%\footnotesize

\bibliographystyle{plain}
\bibliography{ipv6}

\appendix
\newpage

\section{Additional Figures}

\begin{figure}[h!]
    \centering
    \includegraphics[trim=125 0 95 120, clip, width=0.5\textwidth]{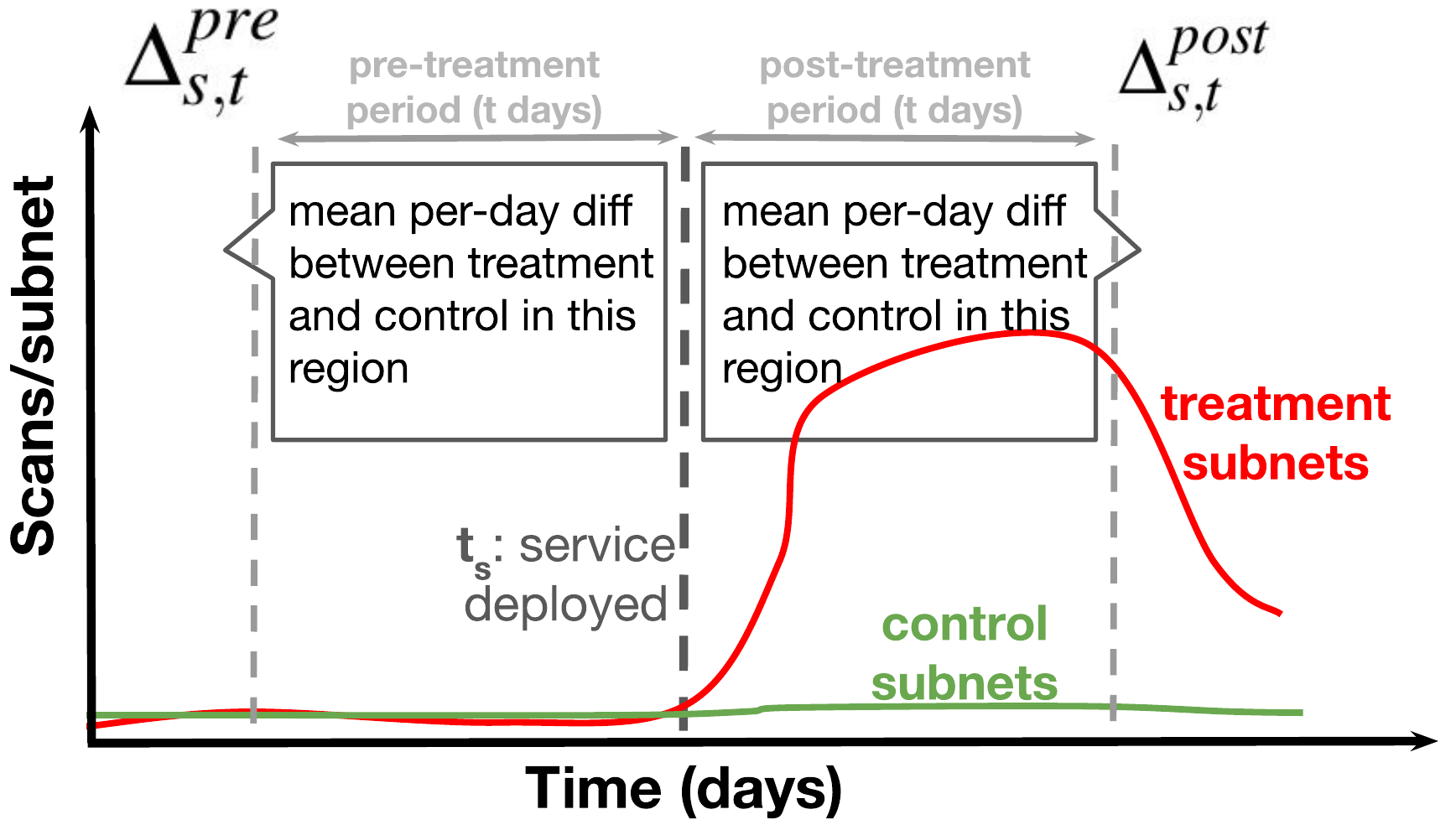}
    \caption{An illustration of the measured $\Delta_{s, t}^{\text{pre}}$ and
    $\Delta_{s, t}^{\text{post}}$ for a service deployed at $t_s$ (as explained
    in \Cref{sec:discovery:methodology:causal}). The corresponding measured
    treatment effect size is \ddiff{s, t} = $\Delta_{s,t}^{post}
    - \Delta_{s,t}^{pre}$.}
    \label{fig:discovery:methodology:effectsize}
\end{figure}

\para{Illustration of the measured effect sizes.}
\Cref{fig:discovery:methodology:effectsize} illustrates the measured effect
sizes from a service deployment inside a \subnet{64} subnet.

\begin{figure}[h!]
  \centering
  \includegraphics[width=.5\textwidth]{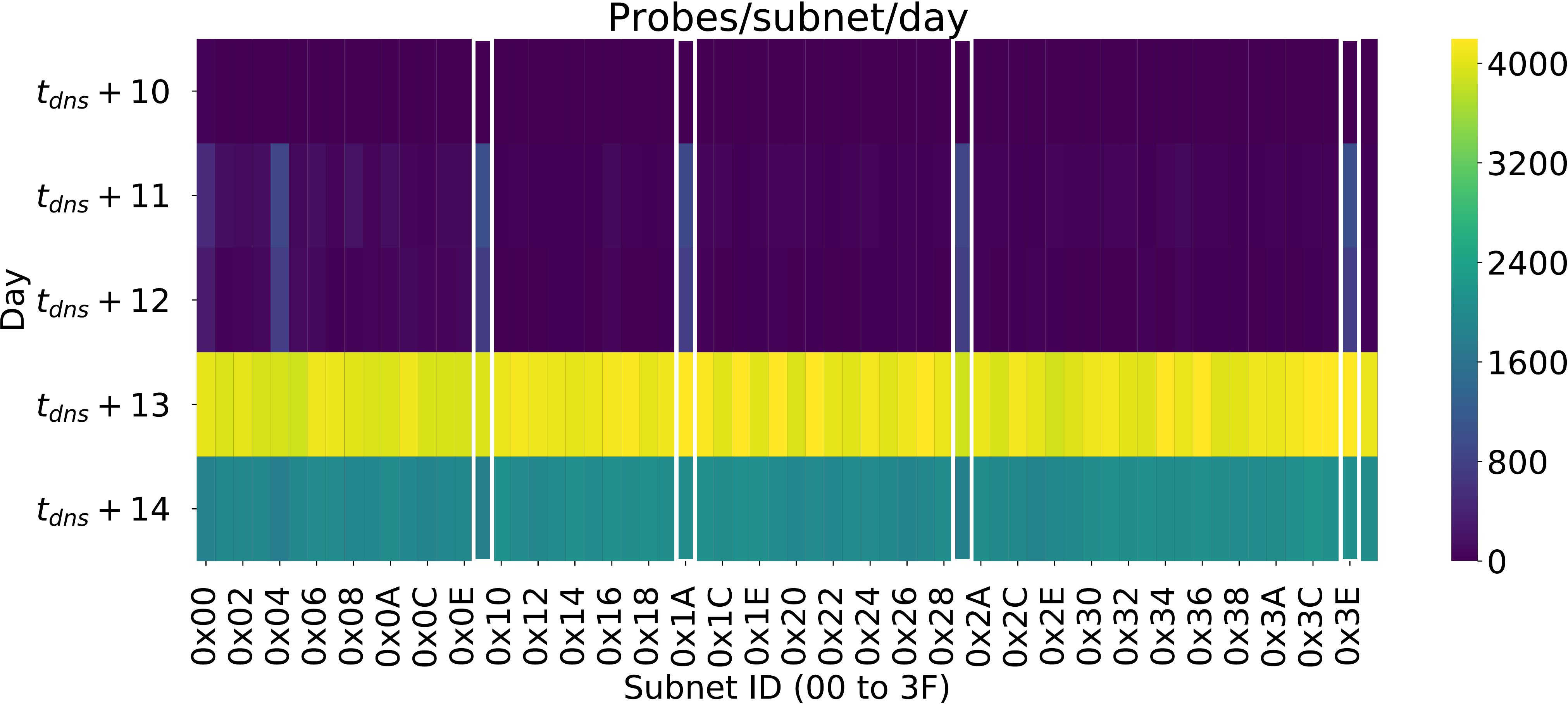}
  \caption{IP probes sent to a \subnet{58} region of our allocation after
  deployment of our DNS probe service (corresponding to the day 10-15 region of
  \Cref{fig:discovery:results:ip-effects:dns}). White highlighted cells
  indicate the subnets used for our DNS probe services (treatment group).}
  \label{fig:discovery:results:ip-effects:backformore-dns}
  %\vspace{-.5em}
\end{figure}

\para{Illustration of effects of direct contact with potential scanners.}
\Cref{fig:discovery:results:ip-effects:backformore-dns} illustrates that scanners arriving during our (direct contact) DNS
probe experiment first probed the treatment region for two days before
returning to scan the addresses located in the lower bytes of all our
\subnet{64} subnets.

\begin{figure*}[t!]
  \centering
  \begin{subfigure}[t]{.33\textwidth}
    \includegraphics[width=\textwidth]{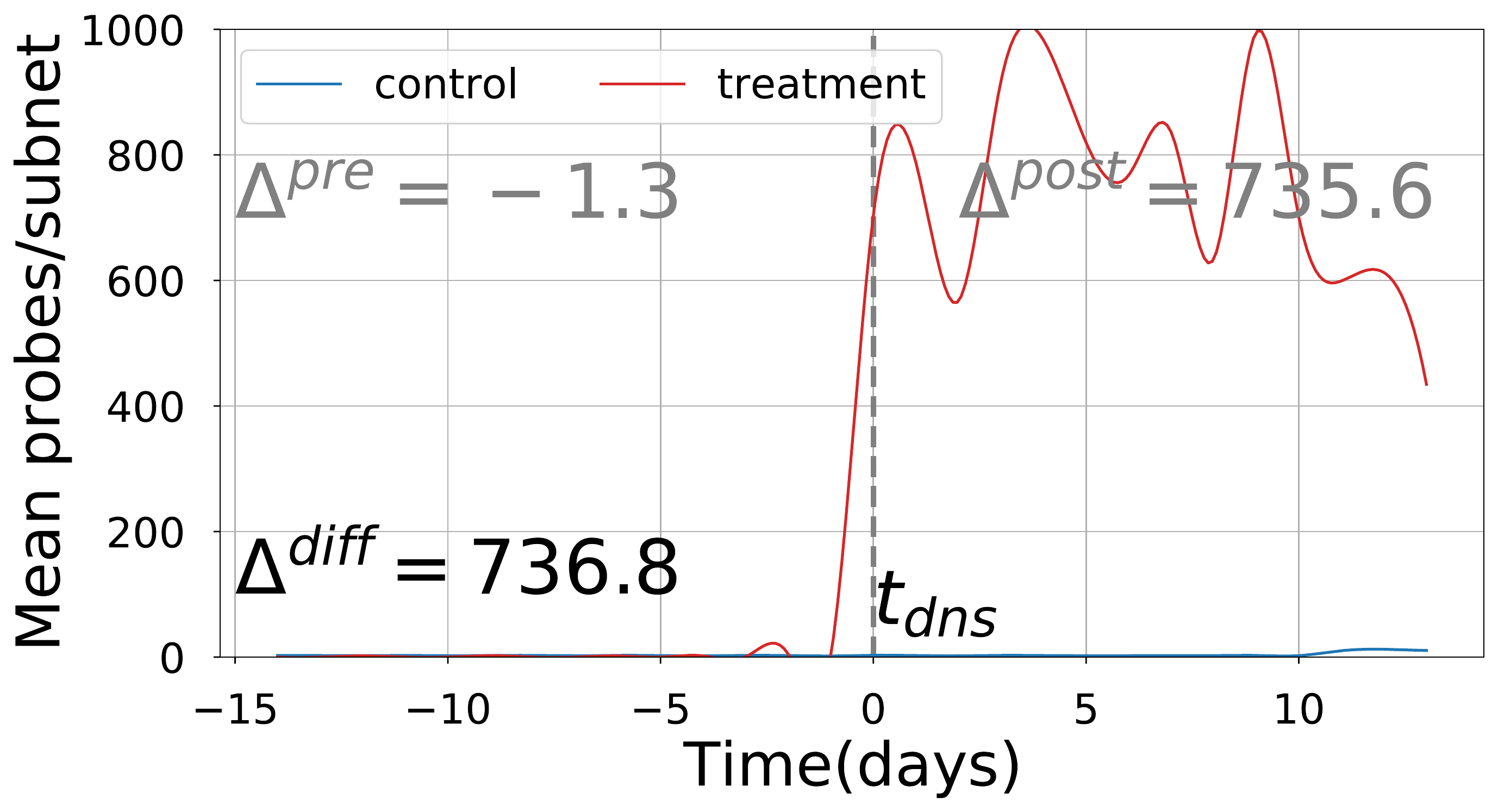}
    \caption{DNS probes}
    \label{fig:discovery:results:dns-effects:dns}
  \end{subfigure}
  \begin{subfigure}[t]{.33\textwidth}
    \includegraphics[width=\textwidth]{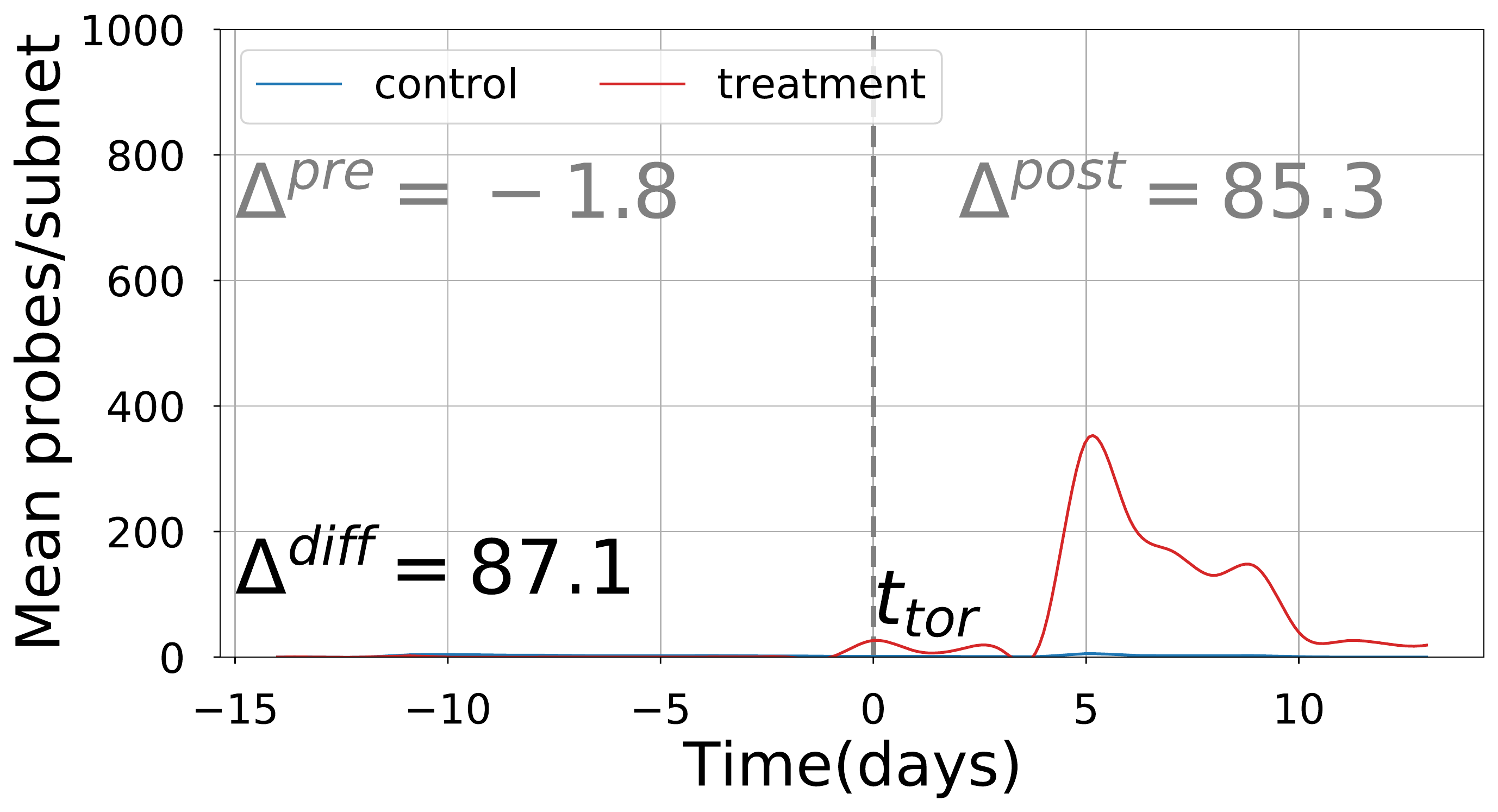}
    \caption{Tor relay}
    \label{fig:discovery:results:dns-effects:tor}
  \end{subfigure}
  \begin{subfigure}[t]{.33\textwidth}
    \includegraphics[width=\textwidth]{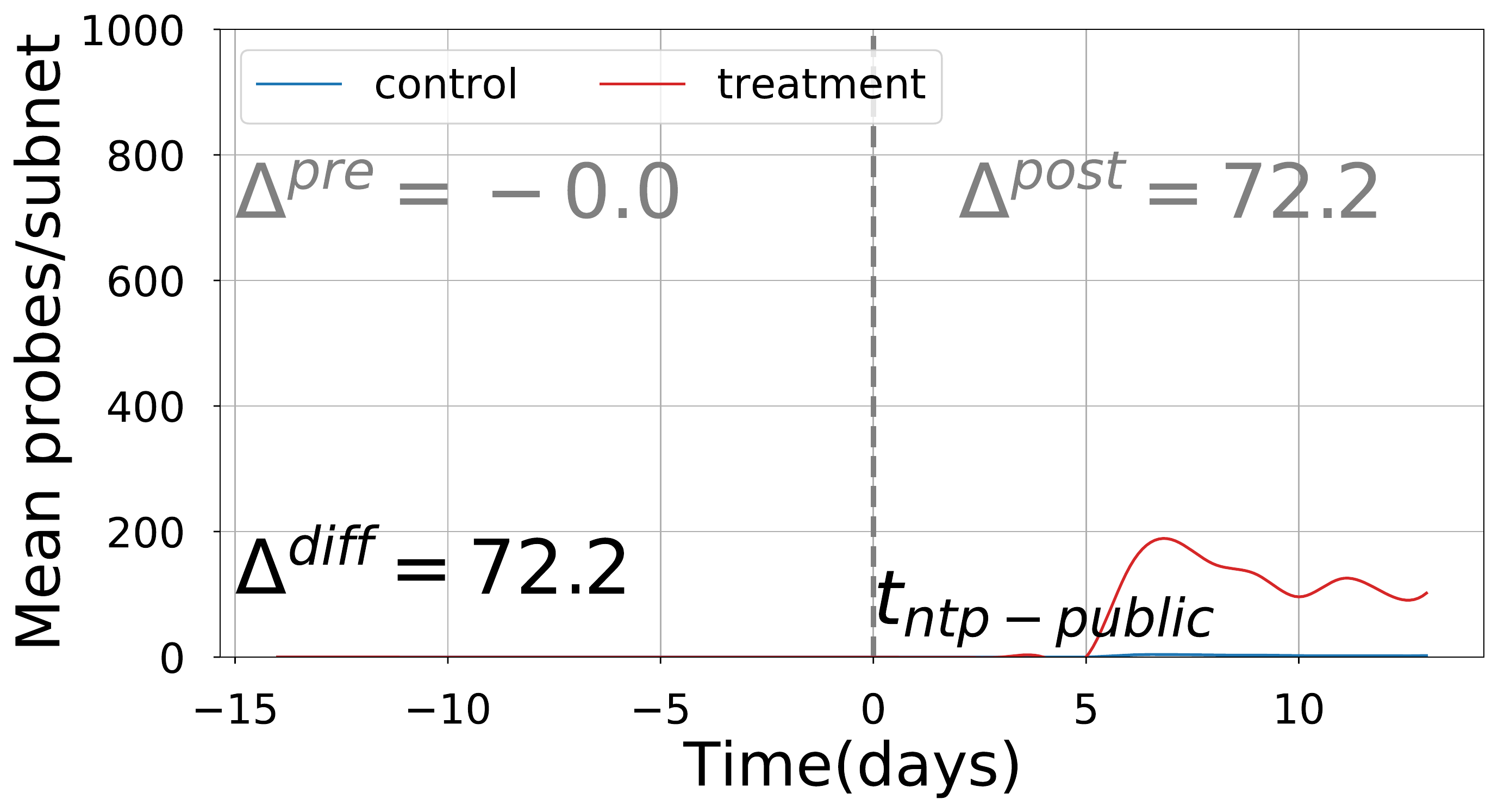}
    \caption{Public NTP server}
    \label{fig:discovery:results:dns-effects:ntp-public}
  \end{subfigure}

  \begin{subfigure}[t]{.33\textwidth}
    \includegraphics[width=\textwidth]{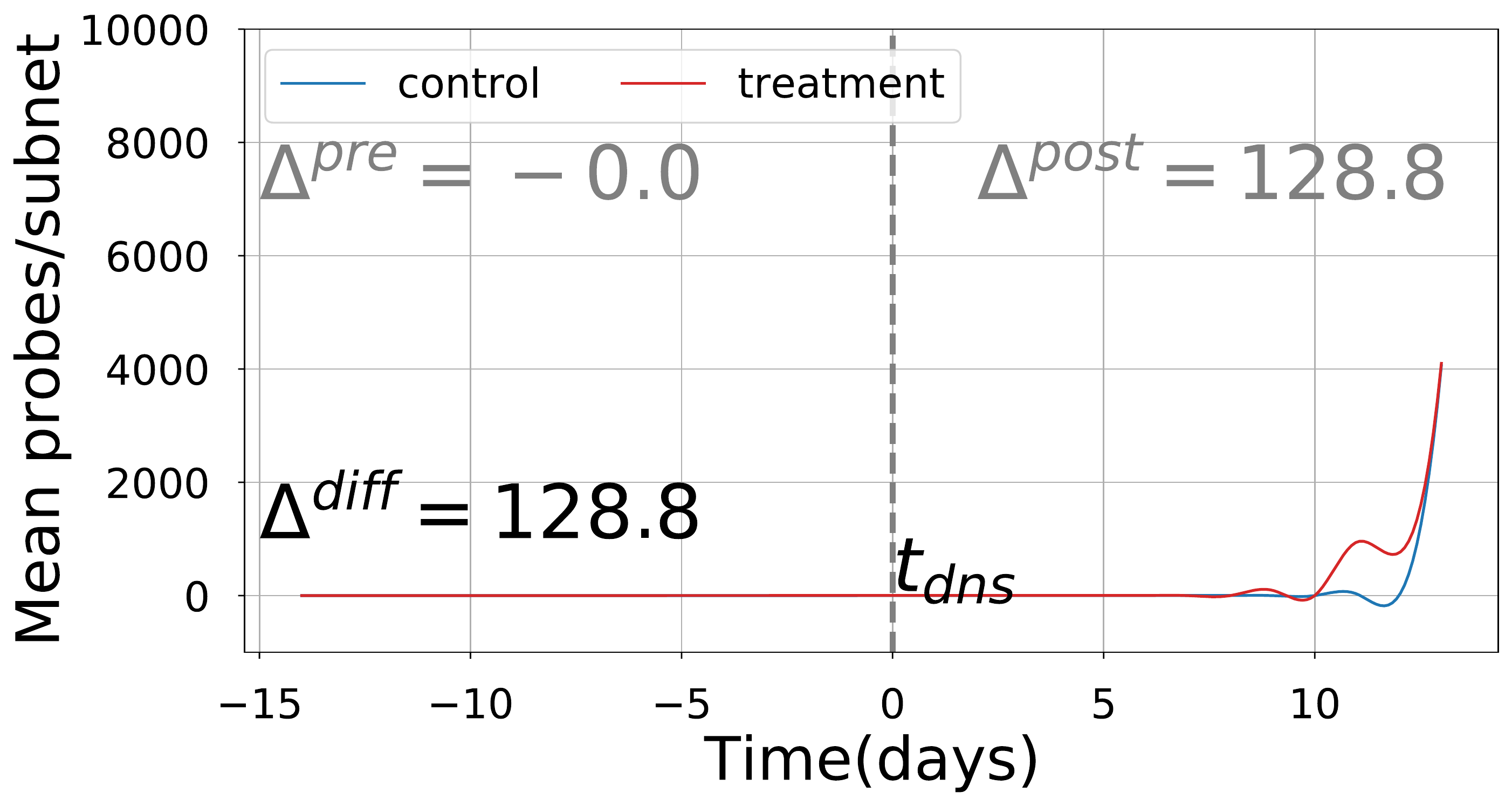}
    \caption{DNS probes}
    \label{fig:discovery:results:ip-effects:dns}
  \end{subfigure}
  \begin{subfigure}[t]{.33\textwidth}
    \includegraphics[width=\textwidth]{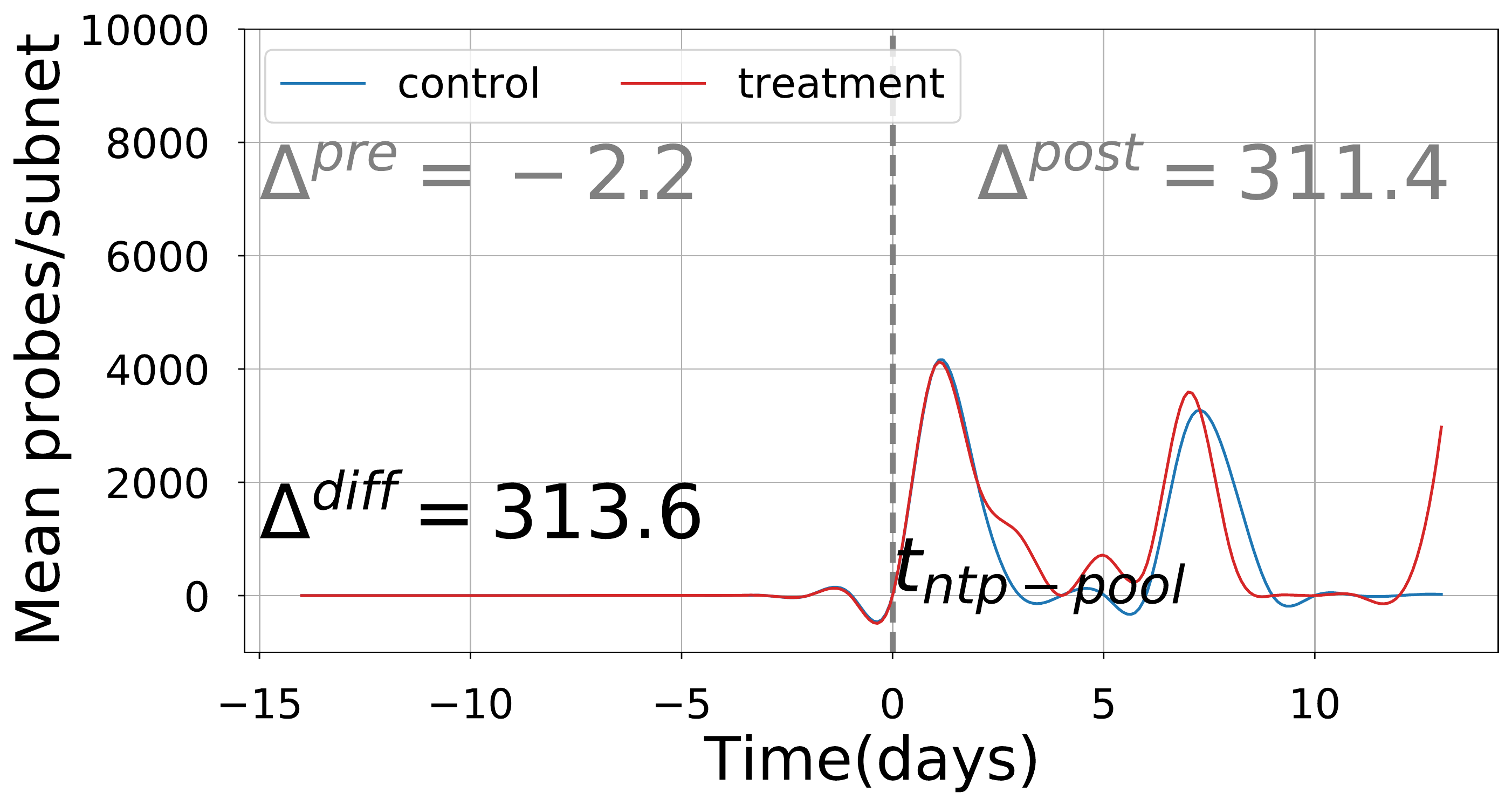}
    \caption{NTP pool server}
    \label{fig:discovery:results:ip-effects:ntp-pool}
  \end{subfigure}
  \begin{subfigure}[t]{.33\textwidth}
    \includegraphics[width=\textwidth]{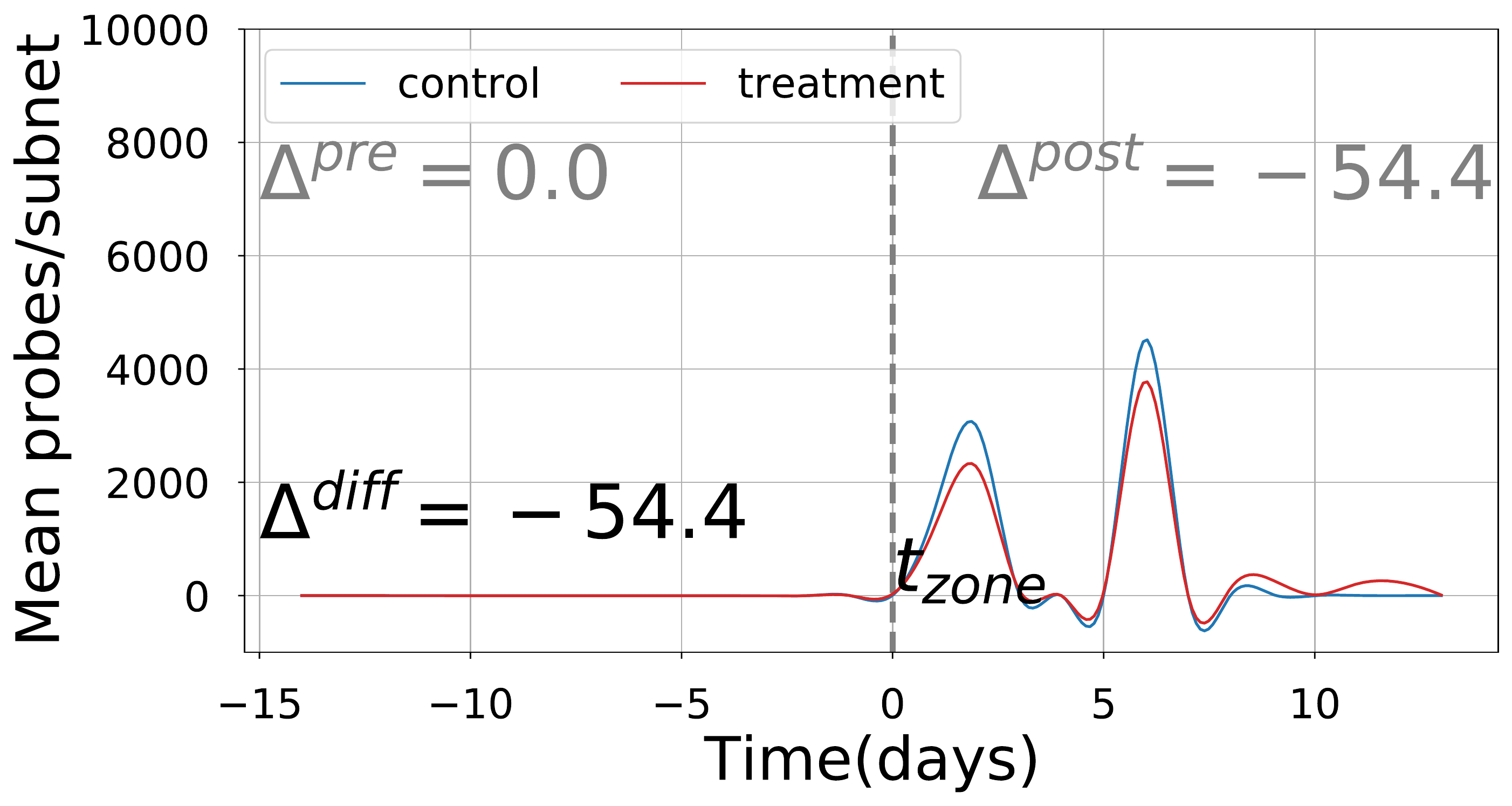}
    \caption{Zone file listing}
    \label{fig:discovery:results:ip-effects:zone}
  \end{subfigure}

  \caption{Treatment effect sizes from deployed services observed in DNS
  scanning logs (\ref{fig:discovery:results:dns-effects:dns},
   \ref{fig:discovery:results:dns-effects:tor}, \ref{fig:discovery:results:dns-effects:ntp-public})
   and PCAPs (\ref{fig:discovery:results:ip-effects:dns}, 
   \ref{fig:discovery:results:ip-effects:ntp-pool}, \ref{fig:discovery:results:ip-effects:zone}). 
   Only statistically significant effect are
  illustrated (Complete results in \Cref{sec:discovery:results:dns-effects} and
  \Cref{sec:discovery:results:ip-effects}). We have applied
  linear interpolation to the visualizations for smoothing.}
  \label{fig:discovery:results:effects}
\end{figure*}

\para{Selected illustrations of treatment effect sizes.}
\Cref{fig:discovery:results:effects} illustrates the measured statistically
significant treatment effect sizes observed from our experiments.

\vspace*{3in}

\end{document}